# Deconvolution of JWST/MIRI Images: Applications to an Active Galactic Nucleus Model and GATOS Observations of NGC 5728

M. T. Leist[1], C. Packham[1,2], D. J. V. Rosario[3], D. A. Hope[4,5], A. Alonso-Herrero[6], E. K. S. Hicks[7], S. Hönig[8], L. Zhang[1], R. Davies[9], T. Díaz-Santos[10,11], O. González-Martín[12], E. Bellocchi[13,14], P. G. Boorman[15,16], F. Combes[17], I. García-Bernete[18], S. García-Burillo[19], B. García-Lorenzo[20,21], H. Haidar[3], K. Ichikawa[22,23], M. Imanishi[2,24], S. M. Jefferies[4], Á. Labiano[25], N. A. Levenson[26], R. Nikutta[27], M. Pereira-Santaella[28], C. Ramos Almeida[20,21], C. Ricci[29,30], D. Rigopoulou[10,18], W. Schaefer[31], M. Stalevski[32,33], M. J. Ward[34], L. Fuller[1], T. Izumi[2,24], D. Rouan[35], and T. Shimizu[9]

[1] Department of Physics and Astronomy, The University of Texas at San Antonio, 1 UTSA Circle, San Antonio, TX 78249-0600, USA; mason.leist@utsa.edu
[2] National Astronomical Observatory of Japan, National Institutes of Natural Sciences (NINS), 2-21-1 Osawa, Mitaka, Tokyo 181-8588, Japan
[3] School of Mathematics, Statistics and Physics, Newcastle University, Newcastle upon Tyne NE1 7RU, UK
[4] Georgia State University, Physics and Astronomy Department, 25 Park Place, Atlanta, GA 30303, USA
[5] Georgia Tech Research Institute, Electro-Optical Systems Laboratory, 925 Dalney Street NW, Atlanta, GA 30318-0834, USA
[6] Centro de Astrobiología (CAB), CSIC-INTA, Camino Bajo del Castillo s/n, E-28692 Villanueva de la Cañada, Madrid, Spain
[7] Department of Physics and Astronomy, University of Alaska Anchorage, Anchorage, AK 99508-4664, USA
[8] School of Physics and Astronomy, University of Southampton, Southampton SO17 1BJ, UK
[9] Max Planck Institut für Extraterrestrische Physik, Giessenbachstrasse 1, D-85748 Garching bei München, Germany
[10] Institute of Astrophysics, Foundation for Research and Technology-Hellas (FORTH), Heraklion 70013, Greece
[11] School of Sciences, European University Cyprus, Diogenes St., Engomi, 1516 Nicosia, Cyprus
[12] Instituto de Radioastronomía y Astrofísica (IRyA), Universidad Nacional Autónoma de México, Antigua Carretera a Pátzcuaro #8701, ExHda. San José de la Huerta, Morelia, Michoacán, C.P. 58089, México
[13] Departmento de Física de la Tierra y Astrofísica, Fac. de CC Físicas, Universidad Complutense de Madrid, E-28040 Madrid, Spain
[14] Instituto de Física de Partículas y del Cosmos IPARCOS, Fac. CC Físicas, Universidad Complutense de Madrid, E-28040 Madrid, Spain
[15] Cahill Center for Astronomy and Astrophysics, California Institute of Technology, Pasadena, CA 91125, USA
[16] Astronomical Institute of the Czech Academy of Sciences, Boční-II 1401, Praha 4, Prague 141 00, Czech Republic
[17] LERMA, Observatoire de Paris, Collége de France, PSL University, CNRS, Sorbonne University, Paris, France
[18] Astrophysics, University of Oxford, DWB, Keble Road, Oxford OX1 3RH, UK
[19] Observatorio de Madrid, OAN-IGN, Alfonso XII, 3, E-28014 Madrid, Spain
[20] Instituto de Astrofísica de Canarias, Calle Vía Láctea, s/n, E-38205 La Laguna, Tenerife, Spain
[21] Departamento de Astrofísica, Universidad de La Laguna, E-38206 La Laguna, Tenerife, Spain
[22] Astronomical Institute, Tohoku University, Aramaki, Aoba-ku, Sendai, Miyagi 980-8578, Japan
[23] Frontier Research Institute for Interdisciplinary Sciences, Tohoku University, Sendai 980-8578, Japan
[24] Department of Astronomy, School of Science, Graduate University for Advanced Studies (SOKENDAI), Mitaka, Tokyo 181-8588, Japan
[25] Telespazio UK for the European Space Agency, ESAC, Camino Bajo del Castillo s/n, E-28692 Villanueva de la Cañada, Spain
[26] Space Telescope Science Institute, 3700 San Martin Drive, Baltimore, MD 21218, USA
[27] NSF's National Optical-Infrared Astronomy Research Laboratory (NOIRLab), 950 N. Cherry Avenue, Tucson, AZ 85719, USA
[28] Instituto de Física Fundamental, CSIC, Calle Serrano 123, E-28006 Madrid, Spain
[29] Núcleo de Astronomía de la Facultad de Ingeniería, Universidad Diego Portales, Av. Ejército Libertador 441, Santiago, Chile
[30] Kavli Institute for Astronomy and Astrophysics, Peking University, Beijing 100871, People's Republic of China
[31] Teaching Learning and Digital Transformation, Academic Innovation, The University of Texas at San Antonio, 1 UTSA Circle, San Antonio, TX 78249-0600, USA
[32] Astronomical Observatory, Volgina 7, 11060 Belgrade, Serbia
[33] Sterrenkundig Observatorium, Universiteit Gent, Krijgslaan 281-S9, Gent B-9000, Belgium
[34] Centre for Extragalactic Astronomy, Department of Physics, Durham University, South Road, Durham DH1 3LE, UK
[35] LESIA, Observatoire de Paris, Université PSL, CNRS, Sorbonne Université, Sorbonne Paris Citeé, 5 place Jules Janssen, F-92195 Meudon, France

## Abstract

The superb image quality, stability, and sensitivity of JWST permit deconvolution techniques to be pursued with a fidelity unavailable to ground-based observations. We present an assessment of several deconvolution approaches to improve image quality and mitigate the effects of the complex JWST point-spread function (PSF). The optimal deconvolution method is determined by using WebbPSF to simulate JWST's complex PSF and MIRISim to simulate multiband JWST/Mid-Infrared Imager Module (MIRIM) observations of a toy model of an active galactic nucleus (AGN). Five different deconvolution algorithms are tested: (1) Kraken deconvolution, (2) Richardson–Lucy, (3) the adaptive imaging deconvolution algorithm, (4) sparse regularization with the Condat–Vũ algorithm, and (5) iterative Wiener filtering and thresholding. We find that Kraken affords the greatest FWHM reduction of the nuclear source of our MIRISim observations for the toy AGN model while retaining good photometric integrity across all simulated wave bands. Applying Kraken to Galactic Activity, Torus, and Outflow Survey (GATOS) multiband JWST/MIRIM observations of the Seyfert 2 galaxy NGC 5728, we find that the algorithm reduces the FWHM of the nuclear source by a factor of 1.6–2.2 across all five filters. Kraken images facilitate detection of extended nuclear emission ∼2″.5 (∼470 pc, position angle ≃ 115°) in the SE–NW direction, especially at the longest wavelengths. We demonstrate that Kraken is a powerful tool to enhance faint features otherwise hidden in the complex JWST PSF.

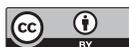

*Unified Astronomy Thesaurus concepts:* Deconvolution (1910); James Webb Space Telescope (2291); Active galactic nuclei (16)

## 1. Introduction

The central engine of an active galactic nucleus (AGN), comprised of a hot and turbulent accretion disk around a central supermassive black hole (SMBH; $M \sim 10^{6-9.5} \, M_\odot$; Kormendy & Richstone 1995) surrounded by a geometrically and optically thick torus of gas and dust, plays a key role in feedback between the AGN, host galaxy, and intergalactic medium (e.g., Kormendy & Ho 2013; Heckman & Best 2014). AGN torus models (e.g., Hönig et al. 2006; Nenkova et al. 2008a, 2008b; Schartmann et al. 2008; Hönig & Kishimoto 2010; Stalevski et al. 2012; Siebenmorgen et al. 2015; Nikutta et al. 2021) reproduced well the mid-infrared (MIR; 7–25 $\mu$m) spectral energy distributions (SEDs) of ground-based 8 m class imaging (e.g., Radomski et al. 2003; Packham et al. 2005; Mason et al. 2006; Radomski et al. 2008; Asmus et al. 2014; Asmus 2019) and spectroscopic observations of local AGNs (e.g., Ramos Almeida et al. 2009; Alonso-Herrero et al. 2011; Ramos Almeida et al. 2011; García-Bernete et al. 2015; Ichikawa et al. 2015; García-González et al. 2017; García-Bernete et al. 2019; González-Martín et al. 2019; García-Bernete et al. 2022) and constrained the torus size to <10 pc radii. High spatial resolution Atacama Large Millimeter/submillimeter Array (ALMA) submillimeter observations have further detected the dusty molecular tori in several nearby Seyfert AGN, suggesting a molecular torus region of up to ∼30 pc radius (depending on the molecular gas tracer used; García-Burillo et al. 2016; Imanishi et al. 2016, 2018; García-Burillo et al. 2019; Alonso-Herrero et al. 2021; García-Burillo et al. 2021).

Recent high spatial resolution MIR interferometric observations of nearby AGNs (i.e., NGC 1068 and Circinus) found dusty polar extensions. These polar dust features, spatially resolved on ∼10 pc scales with the Very Large Telescope Interferometer (VLTI)/Mid-Infrared Interferometric Instrument (MIDI; e.g., Burtscher et al. 2013; Hönig et al. 2013; López-Gonzaga et al. 2014; Tristram et al. 2014; López-Gonzaga et al. 2016; Leftley et al. 2018) and VLTI/Multi AperTure mid-Infrared SpectroScopic Experiment (MATISSE; e.g., Gámez Rosas et al. 2022; Isbell et al. 2022), show that polar dust accounts for ∼50%–80% of the 8–13 $\mu$m emission, complicating the interpretation of the MIR emission source at these scales. The polar dust is found at a similar position angle (PA) as the much larger-scale (few ∼100 pc) extended MIR emission detected in 8 m class imaging (e.g., Asmus et al. 2016; García-Bernete et al. 2016; Asmus 2019; Alonso-Herrero et al. 2021) and, in most cases, found perpendicular to the ALMA-identified dusty molecular tori (e.g., García-Burillo et al. 2021), prompting the inclusion of a polar component in models (e.g., Hönig & Kishimoto 2017; Stalevski et al. 2017). However, numerous questions remain about polar dust, such as its physical properties (e.g., grain size distribution, temperature, mass and density distributions, composition, etc.; Ramos Almeida & Ricci 2017; Lyu & Rieke 2018; Hönig 2019; Tazaki & Ichikawa 2020), origin, and its relationship to the ionization cone.

The Galactic Activity, Torus, and Outflow Survey (GATOS[36]) was awarded JWST General Observers (GO) time during Cycle 1 (ID: 2064; PI: D. Rosario) to characterize polar dust in a sample of eight nearby Seyfert galaxies, all with prior evidence for extended dust emission from ground-based MIR imaging (Asmus et al. 2016; Asmus 2019), using multiband imaging obtained with the Mid-Infrared Instrument (MIRI) Imager Module (MIRIM; Bouchet et al. 2015; Rieke et al. 2015a, 2015b; Wright et al. 2015). JWST affords an unprecedented chance to observe such extended dust emission thanks to its vastly superior sensitivity, low background, and very stable image quality. JWST can yield point-spread functions (PSFs) limited by the diffraction limit (i.e., PSF FWHM ∼ $\lambda/D$ at wavelengths > 2 $\mu$m) while MIRI achieves Nyquist sampling at wavelengths > 7 $\mu$m, with undersampled PSFs below these wavelengths (see Rigby et al. 2023, for more details). This stable image quality has two major benefits compared to 8 m class ground-based observations: (1) an improved overall PSF for all observations, and (2) the chance to deconvolve the complex, but crucially stable PSF from the observational data.

Deconvolution has served as a powerful tool for signal and image processing in diverse fields such as seismology (e.g., Gal et al. 2016), medical imaging (e.g., Shajkofci & Liebling 2020), and astronomy (e.g., Farrens et al. 2017), affording a reliable recovery and accurate estimate of the true source properties before convolution with the PSF inherent the data collecting and optical system (see Starck et al. 2002, for a review). Every astronomical instrument has a unique PSF, characterized as a 2D representation of the instrument's response to a point source. Mathematically, this can be described at the coordinates $(x, y)$ by the convolution equation (Bracewell & Roberts 1954):

$$I(x, y) = (O * P)(x, y) + N(x, y), \quad (1)$$

where $I(x, y)$ is the calibrated observed image, $O(x, y)$ is the "truth" image before convolution, $P(x, y)$ is the PSF of the instrument, $N(x, y)$ is the noise introduced during the observation (e.g., detector noise, optical system noise, sky background noise, etc.), and $*$ denotes convolution. The goal of deconvolution is to determine $O(x, y)$ from known observation $I(x, y)$ and PSF $P(x, y)$ (Starck & Murtagh 2002). This is often referred to as an "ill-posed problem," as a unique solution is typically impossible.

Images obtained with optical/infrared (IR) ground-based telescopes suffer degraded image quality due to external (atmospheric turbulence and scattering, water vapor, etc.) and internal (imperfect mirrors, alignment errors, diffraction, scattering, etc.) factors along the optical path. Deconvolution techniques have leveraged advances in computational power and ameliorated image quality and stability on telescopes (i.e., through adaptive optics (AO) systems) to reduce degradations and improve the delivered image (Starck & Murtagh 2002). For example, Richardson–Lucy–deconvolved MIR Keck Telescope images revealed extended emission associated with the narrow emission line region of the AGN in NGC 1068 (Bock et al. 2000) and the ionization cone of Cygnus A (Radomski et al. 2002).

---

[36] https://gatos.myportfolio.com/

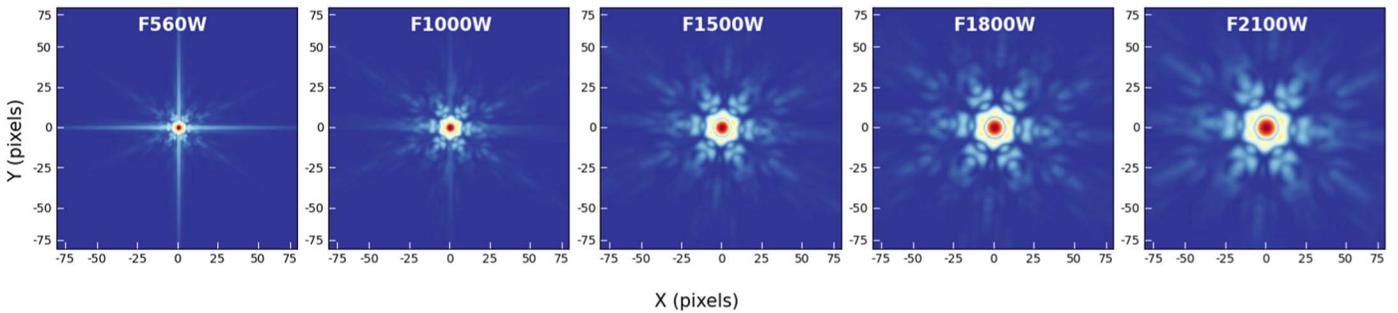

**Figure 1.** Log-scaled detector-sampled PSF images used for deconvolution generated by WebbPSF, showing the central 150 × 150 pixels to focus on the details of the PSF, in the filters as noted.

Deconvolution can be complicated in astronomy as many images contain both smooth and sharp features, such as extended objects and point sources (e.g., extended galaxies and stars). The algorithm converges to an optimal solution after different numbers of iterations (i.e., point sources converge faster than extended objects) causing one feature to "overdeconvolve" and the other feature to "underdeconvolve." This can cause "ringing artifacts," manifested as a series of concentric rings, most clearly defined around point sources (i.e., high spatial frequency; see Magain et al. 1998, for more details). An additional artifact that is often introduced during deconvolution is that of a mottled effect in smooth background objects (i.e., low spatial frequency), sometimes referred to as the "orange peel" effect. It is therefore crucial when considering the methodology to find a balance between data fidelity (i.e., conservation of flux), smoothness (i.e., ringing artifact suppression), and improved image quality of the restored image. Deconvolution algorithms provide this balance by employing constraints (known as regularization criteria) to recover an image that is closest to the truth image $O(x, y)$ (see Starck et al. 2002, for a review). Commonly used regularization criteria include convergence criteria (i.e., an image quality measurement), noise statistics (i.e., ensuring an approximately consistent noise, to a user-defined value, between the initial and final iteration), positivity of the resultant image (i.e., nonnegative pixel values), and residual image statistics (i.e., difference between the original and resultant image; McNeil & Moody 2005).

In this paper, we present our deconvolution testing methodology to find the optimal balance of data fidelity and image quality improvement when restoring (1) simulated JWST/MIRIM observations of a toy AGN model through several filters and (2) GATOS JWST/MIRIM Cycle 1 observations of the Seyfert 2 galaxy NGC 5728. The paper is organized as follows: in Section 2 we present a description of our JWST PSF modeling, and simulate MIRIM images of a toy AGN model and JWST/MIRIM observations of NGC 5728. Section 3 details the deconvolution algorithms tested, and our deconvolution convergence criteria. Section 4 compares the deconvolution results of the toy AGN model for each deconvolution algorithm. Section 5 discusses the results of applying deconvolution to JWST/MIRI imaging. The conclusions are summarized in Section 6.

## 2. Simulated and Observed Images

### 2.1. Point-spread Function Modeling

A key input to the deconvolution methods are the JWST and JWST/MIRIM PSFs for each of the five filters used. JWST's characteristic hexagonal primary mirror segments create a distinctive diffraction pattern comprising of a high Strehl ratio core, six bright diffraction spikes, a hexagonal Airy disk, and a complex but fainter extended structure at distances further from the Airy disk (i.e., Rigby et al. 2023). However, while complex, the PSF is very stable and well characterized.

To model the JWST PSF, we utilize WebbPSF (version 1.1.1; Perrin et al. 2012), a Python-based package that simulates the PSF for four JWST instrument observing modes. WebbPSF transforms optical path difference maps of the telescope and each instrument into PSFs, accounting for detector pixel scales, rotations, filter profiles, and the input point source's spectra.[37] Wave front measurements are made regularly and made available in the Mikulski Archive for Space Telescopes (MAST),[38] which can be updated within WebbPSF to produce time-dependent PSF models (Rigby et al. 2023). We generated PSFs for the five filters below (Section 2.3) using WebbPSF sampled to the MIRIM detector, centered on the SUB256 subarray. The WebbPSF PSFs do not include contributions from the MIRI detector, such as read noise, bad pixels, cosmic rays, etc. but do include the "crosshair" effect. The crosshair effect was first found in the Spitzer Space Telescope's IRAC Si:As detectors (Pipher et al. 2004) and is an instrumental effect in the MIRIM x- and y-coordinate frame, most obviously at wavelengths ⩽ 10 μm, attributed to internal diffraction within the detector's electrical contacts (Gáspár et al. 2021). The WebbPSF PSFs were used as reference PSFs during deconvolution (see Sections 4 and 5; Figure 1).

### 2.2. Toy AGN Model

To simulate the effectiveness and fidelity of deconvolution methods for AGN research, we created a "toy model." In this model the AGN consists of four key thermally emissive components: (1) the central region (dominantly <20 pc diameter MIR emission from the obscuring torus illuminated by the accretion disk of the central engine), (2) an elongated dusty polar extension, extending ∼250 pc from the central region, (3) a dusty ionization bicone extending hundreds of pc of parsecs from the central region, and (4) a kiloparsec-scale host galaxy. Using the JWST/MIRIM plate scale (1 pixel = 0.″11; Bouchet et al. 2015) at the distance to NGC 5728 (39 Mpc; Rest et al. 2014) we estimate the physical scale for each model component (1″ ∼ 190 pc).

Even with JWST's diffraction-limited high-sensitivity observations, the <20 pc MIR-emitting central region component (i.e., smaller than the molecular torus diameter) will be unresolved at the distance of NGC 5728, which instead will appear as a bright point source (Figure 2). We therefore model this component as a PSF for the five filters below (Section 2.3) using WebbPSF sampled to the MIRI detector, centered on the SUB256 (hereafter central region PSF; Section 2.1).

---

[37] See the WebbPSF user documentation for more details: https://webbpsf.readthedocs.io/.
[38] https://mast.stsci.edu/portal/Mashup/Clients/Mast.portal.html

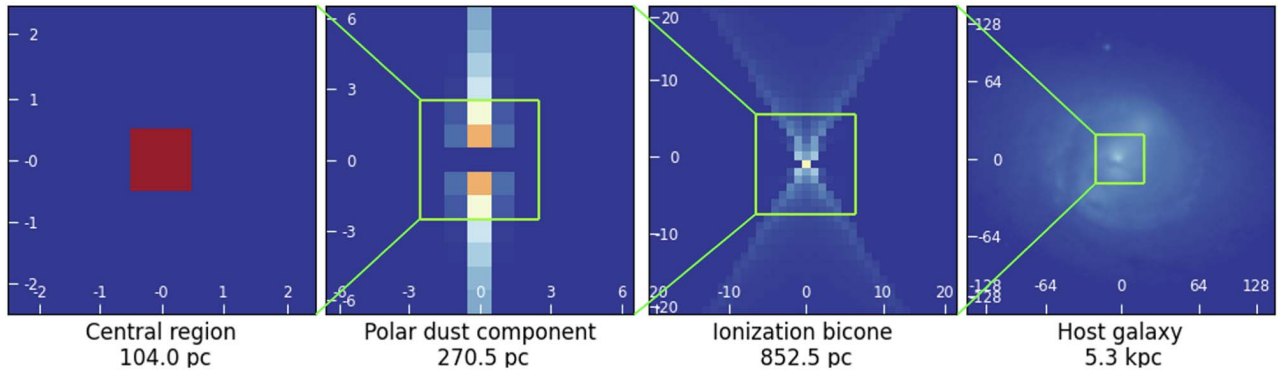

**Figure 2.** Toy AGN model components, from left to right: the central region (dominantly the obscuring torus illuminated by the accretion disk of the central engine), the polar dust component, the ionization bicone, and the host galaxy. Each figure is log scaled (except for the central region, which is linear scaled), scaled to the central region component. The *x*- and *y*-axes are given in pixels, and the size of the field of view (FOV) for each region is given at parsec scales below each frame (1 pixel = 20.8 pc).

The polar dust feature of NGC 1068 (Gámez Rosas et al. 2022) is similar to a compact cone which augments the flux in the ionization bicone, where the flux is concentrated to ≲10 pc. Like the central region component, such a feature will be unresolved at the distance of NGC 5728. However, some radiative hydrodynamical models (Williamson et al. 2020) result in a collimated column of polar-wind–driven thermal dust emission extending from the central engine to >100 pc. MIR linear extensions have been observed in nearby AGNs, such as the ∼40 pc bar-like extension from the nucleus of the Circinus galaxy (e.g., Packham et al. 2005), interpreted by the radiative transfer simulations of Stalevski et al. (2017). This work employs a phenomenological dust emission model consisting of a compact dusty disk and hollow cone, whose interior is illuminated by a tilted accretion disk, giving rise to one arm of the bicone being substantially brighter than the other. To explore the extent to which the detection of such features could be aided by deconvolution, we model a linear feature at a position distinct from the ionization cone: we term this the polar dust component hereafter. The counts per pixel decrease as $r^{-1}$ from the central region (central PA = 0° with respect to the image array), the outer counts per pixel decreases as $r^{-3}$ perpendicular to the outside faces, and the central pixel is set to zero to account for dust sublimation (i.e., Barvainis 1987, Figure 2). 95% of the counts of this component are contained within an aperture of 1″21 (i.e., ⩽228.8 pc from the central region).

NGC 5728 has a biconical ionization structure extending ∼1.7 kpc to the SE (PA = 118°; Schommer et al. 1988; Arribas & Mediavilla 1993; Wilson et al. 1993; Mediavilla & Arribas 1995; Shimizu et al. 2019) and ∼2.1 kpc to the NW (PA = 304°; Wilson et al. 1993; Shimizu et al. 2019) of the nucleus, with similar opening angles between the two cones ∼55°–65° (Wilson et al. 1993; Shimizu et al. 2019). An AGN-driven weak outflow within the bicone is detected 130 pc to the SE and 230 pc to the NW (Shimizu et al. 2019). We model the ionization structure as an edge-brightened bicone consisting of two axisymmetric cones with a shared apex. Each cone has the same (1) opening angle of 60°, (2) inclination angle of 90° (i.e., in the plane of the sky), and (3) central PA of 0° (with respect to the image array). The counts per pixel along the cone arms decrease as $r^{-2}$ from the apex, the cone's inner face counts per pixel decrease as $r^{-1}$ perpendicular to the inner cone face, and the cone's outer counts per pixel decrease as $r^{-3}$ perpendicular to the outside face of the cone (Figure 2). The counts were set to zero at the apex of the bicone due to sublimation from the central engine. These model parameters produce an acceptable representation of an edge-brightened, partly filled ionization bicone. 95% of the counts of the bicone are contained within an aperture of 4″07 (⩽769.6 pc).

The final component of the toy AGN model is the host galaxy, for which we use NGC 5728. This object has a large-scale stellar bar ∼11 kpc (PA = 33°; Schommer et al. 1988; Prada & Gutiérrez 1999) and an ∼800 pc nuclear stellar bar (PA = 85°; Shaw et al. 1993) surrounded by a circumnuclear star formation ring (Schommer et al. 1988; Shaw et al. 1993; Wilson et al. 1993; Capetti et al. 1996; Prada & Gutiérrez 1999). We used an archival near-infrared Hubble Space Telescope/WFC3 F110W (PI: J. Greene, ID: 13755)[39] image of the circumnuclear region of NGC 5728, and extracted the central ∼5.3 × 5.3 kpc (∼28″2 × 28″2 = 256 × 256 pixels, Figure 2), a much larger spatial extent than the other model components. We rotate the extracted image such that the PA of the nuclear stellar bar is rotated by 33° relative to the bicone (i.e., nuclear stellar bar PA = 327° with respect to the image array) matching the ∼33° PA offset of the nuclear stellar bar relative to the SE ionization cone of NGC 5728.

Table 1 lists the physical scales and relative integrated counts of the components compared to the central region PSF. We set the total integrated counts of the central region PSF compared to the polar dust component to 10:1, the bicone to 50:1, and the host galaxy to 200:1. The integrated counts represent a cartoon of AGN components and ensure a high contrast between the components. These count ratios were fixed in all simulated filters to ensure a constant contrast to enable an assessment of each deconvolution algorithm's performance at recovering individual model components across multiband simulated imaging. The polar dust, ionization bicone, and galaxy model components were then coadded to form a three-component model (Figure 3) used as input in our simulations of MIRIM data. Additional input models varying the bicone's total integrated counts and opening angle are discussed in Appendix B.

---

[39] https://hla.stsci.edu/

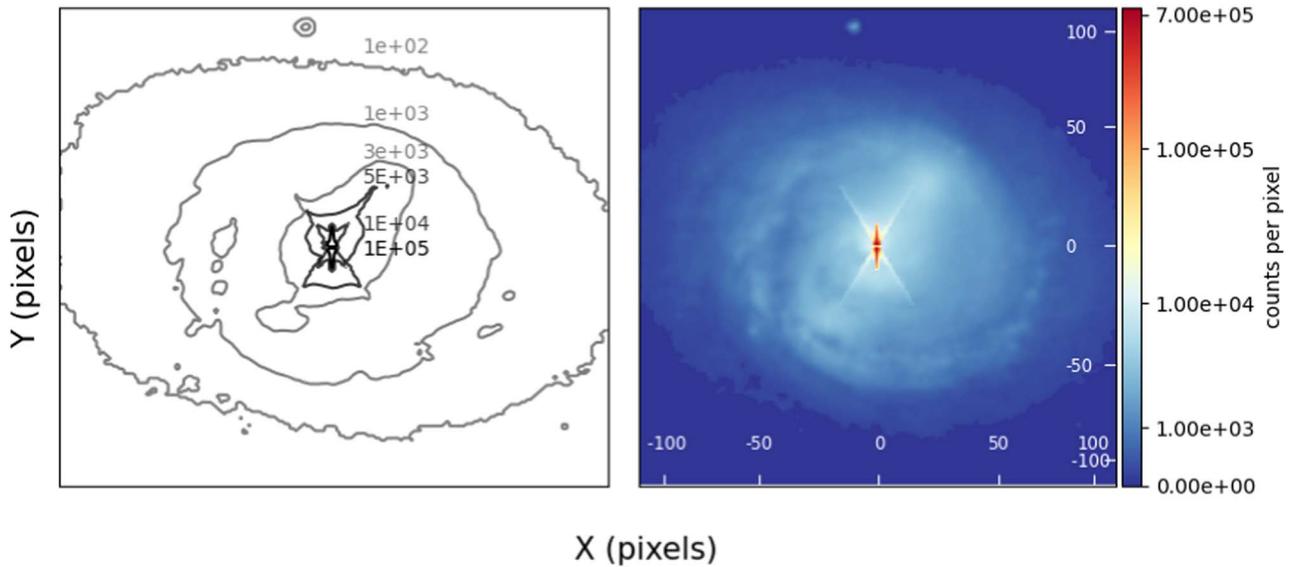

**Figure 3.** Three-component model contour map (left). Contour levels correspond to counts ranging from $10^2$–$10^5$ counts per pixel. Three-component model (right). Image is displayed log scaled, showing the central 200 × 200 pixels (∼22″.2, ∼4.2 kpc). Note the central region component is excluded from this model.

**Table 1**
Model Component Physical Scales and Relative Integrated Intensities

| Model Component | Physical Scale | | | Relative Integrated Intensity |
|---|---|---|---|---|
| | (pixels) | (arcsec) | (pc) | Component: Central Region |
| Central region | 1.0 | 0.11 | 20.8 | ⋯ |
| Polar dust | 13.0 | 1.21 | 228.8 | 10:1 |
| Ionization bicone | 37.0 | 4.07 | 769.6 | 50:1 |
| Host galaxy | 256.0 | 28.2 | 5324.8 | 200:1 |

**Note.** The physical scales for the polar dust and ionization bicone components are where 95% of the counts are enclosed.

We used MIRISim (version 2.4.2; Klaassen et al. 2021), a Python-based simulation package for JWST/MIRI, to generate simulated MIRIM images. MIRISim accepts inputs such as positions, fluxes, images, and observing modes (including subarray, dither pattern, etc.) to simulate images from MIRI.[40] We simulated MIRIM images for each filter listed below using the three-component model following a similar observational setup as used for observations of NGC 5728 (Section 2.3) but with the detector read noise set to zero (in preparation for subsequent photometric scaling), and used only one group (equivalent to a "frame," to avoid subsequent coaddition effects). Next the central region PSF was added to the center of the MIRISim output, then the combined image was flux calibrated to the aperture photometric measurements of NGC 5728 (Table 4) in the respective filters. Finally, a simulated MIRIM read-noise image was generated following the same observational setup as above, flux calibrated, then added to the photometrically scaled image. Pixels with a negative value were set to zero, as some of the deconvolution algorithms make this adjustment as an initial step. Thus high signal-to-noise ratio (S/N) simulated JWST/MIRIM toy AGN model images in five filters were produced. Figure 4 illustrates our methodology to produce the simulated MIRIM images.

### 2.3. JWST/MIRIM Observations

We utilized JWST/MIRIM observations of the Seyfert 2 galaxy NGC 5728, observed on 2023 March 3 as part of the GATOS Cycle 1 GO observing campaign (PI: D. Rosario; ID: 2064), for continued deconvolution experimentation. Images were taken in five MIRIM filters (F560W, F1000W, F1500W, F1800W, and F2100W) using the SUB256 subarray (256 × 256 pixels, ∼28″.2 × 28″.2) with the FASTR1 detector readout mode (see D. J. V. Rosario et al. 2024, in preparation for more details). We obtained raw data directly from the MAST archive and processed it through the JWST pipeline (version 10.2; Bushouse et al. 2023) and introduced refinements to the pipeline to enhance data quality, discussed in detail in D. J. V. Rosario et al. (2024, in preparation). The primary adjustment pertained to absolute astrometry, which we corrected to match the nuclear position as derived from very long baseline interferometric measurements (R. A. = 14:42:23.872, decl. = 17:15:11.016; Shimizu et al. 2019). For each JWST/MIRIM observation, the centroid of the nuclear source was determined using a quadratic centroiding algorithm[41] in a 45 pixel (5″) box centered on the nucleus (at a resolution of one pixel), and then a 256 × 256 subset of each image was produced.

---

[40] See the MIRISim user documentation for more details: https://wiki.miricle.org/pub/Public/MIRISimPublicRelease2dot3/MIRISim.pdf.

[41] https://photutils.readthedocs.io/en/stable/api/photutils.centroids.centroid_quadratic.html

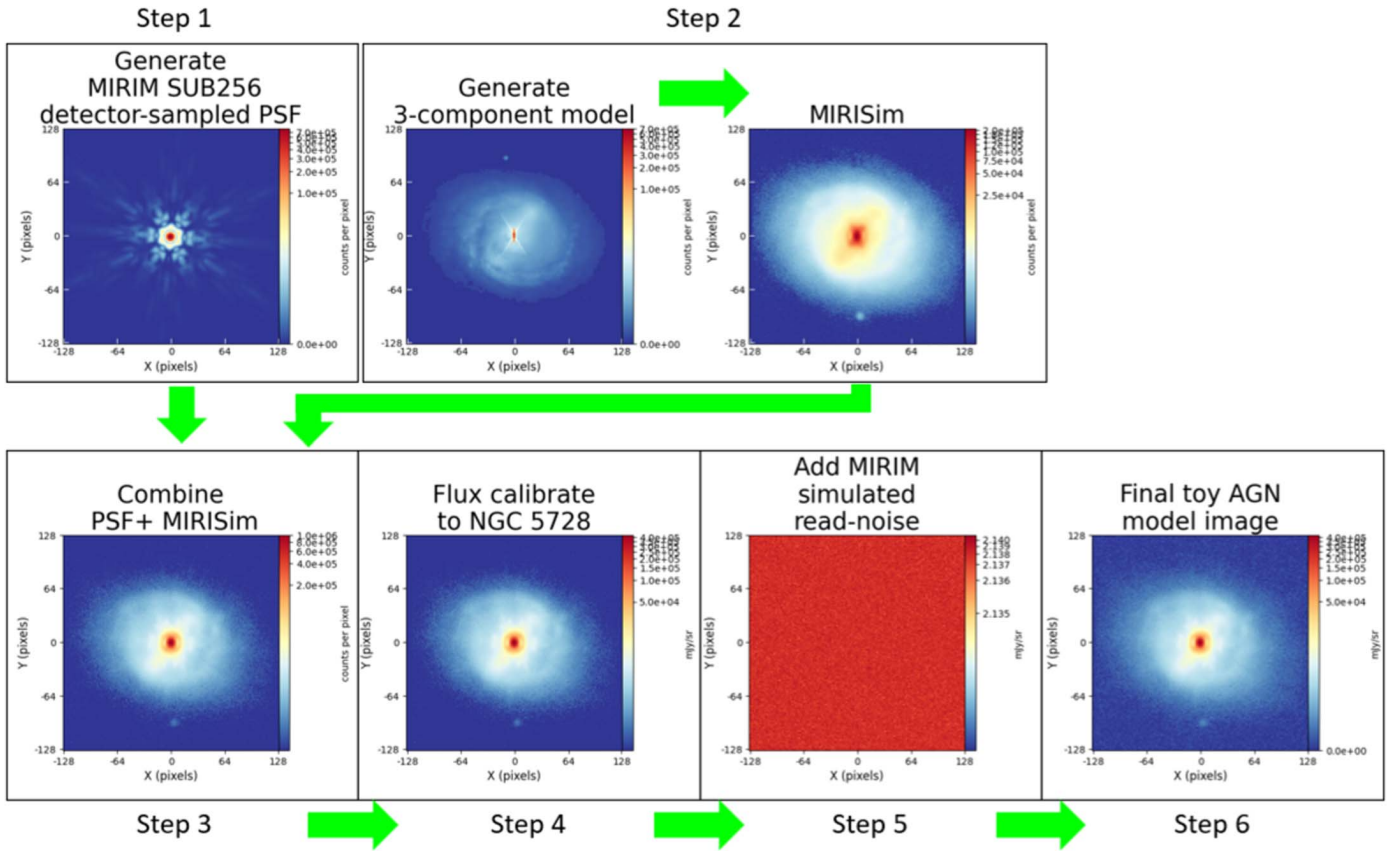

**Figure 4.** Toy AGN model generation flow chart for the F2100W image. Note that this procedure was followed for all five filter images.

## 3. Deconvolution Methodology

### 3.1. The Deconvolution Methods

We tested five iterative deconvolution algorithms which offered different ways of solving the ill-posed problem of deconvolution to recover a best estimate of the truth image $O(x, y)$ (the toy AGN model in our case). The algorithms employed were (1) Kraken deconvolution (Hope et al. 2022), (2) linear regularization using noncirculant Richardson–Lucy[42] (Richardson 1972; Lucy 1974; Ingaramo et al. 2014), (3) myopic deconvolution using the adaptive imaging deconvolution algorithm[43] (AIDA; Hom et al. 2007), (4) sparse regularization (Starck et al. 2015b) with the Condat–Vũ (Vu 2011; Condat 2013) algorithm[44] (SCV; Farrens et al. 2017), and (5) iterative Wiener filtering and thresholding (IWFT; Šroubek et al. 2019). The input parameters used for each deconvolution algorithm are listed in Appendix A, and brief comments on each follow.

Kraken is based on the compact multiframe blind deconvolution (CMFBD) algorithm and designed for processing extreme AO high-cadence imaging (Hope et al. 2022). Kraken works by first computing precise initial guesses of the object and PSF by means of the CMFBD estimator (Hope & Jefferies 2011; Hope et al. 2019) then, from these initial estimates, performs a complete multiframe blind deconvolution (MFBD) procedure (Jefferies & Christou 1993; Schulz 1993) on each image frame before combining to produce a final, high-resolution deconvolved image. The MFBD performance is typically improved if the initial CMFBD estimate for the object has both high-quality and low spatial frequency data (Hope et al. 2016). For our work, with a priori knowledge of the PSFs (i.e., nonblind), Kraken directly estimates the truth image using a Fourier band-limited representation of the object and a nonnegativity constraint on the flux of the estimate. The Fourier band limit is most effective for the F2100W and F1800W images due to the image's lack of measured Fourier components. As more spatial frequencies are measured at shorter wavelengths, the band-limit restriction is relaxed, and only the nonnegativity constraint remains. This flexibility in the object model enables Kraken to estimate high and low spatial frequencies of objects in the image array simultaneously, making it an ideal algorithm for simultaneously deconvolving bright and diffuse emission found in AGNs and their host galaxies.

Richardson–Lucy is a robust linear regularization iterative method widely used for image deconvolution in astronomy and other sciences. Richardson–Lucy has the advantages of (1) each iteration result is nonnegative and (2) flux is conserved both globally and locally in the image array if the background noise is >0. However, there are well-known drawbacks to Richardson–Lucy, namely (1) noise amplification and (2) ringing artifact structures around high spatial frequency features and image array edges, both increasing in severity with the iteration number (Magain et al. 1998). Several variations of Richardson–Lucy have been developed to mitigate these effects (i.e., total variation (TV) regularization (e.g., Dey et al. 2006), vector acceleration (e.g., Biggs & Andrews 1997), and noncirculant edge handling[45] (e.g.,

---
[42] https://github.com/clij/clij2-fft
[43] https://github.com/erikhom/aida
[44] https://github.com/CEA-COSMIC/pysap-astro

[45] First introduced in the 2014 Grand Challenge on Deconvolution: http://bigwww.epfl.ch/deconvolution/challenge2013/index.html?p=doc_math_rl.

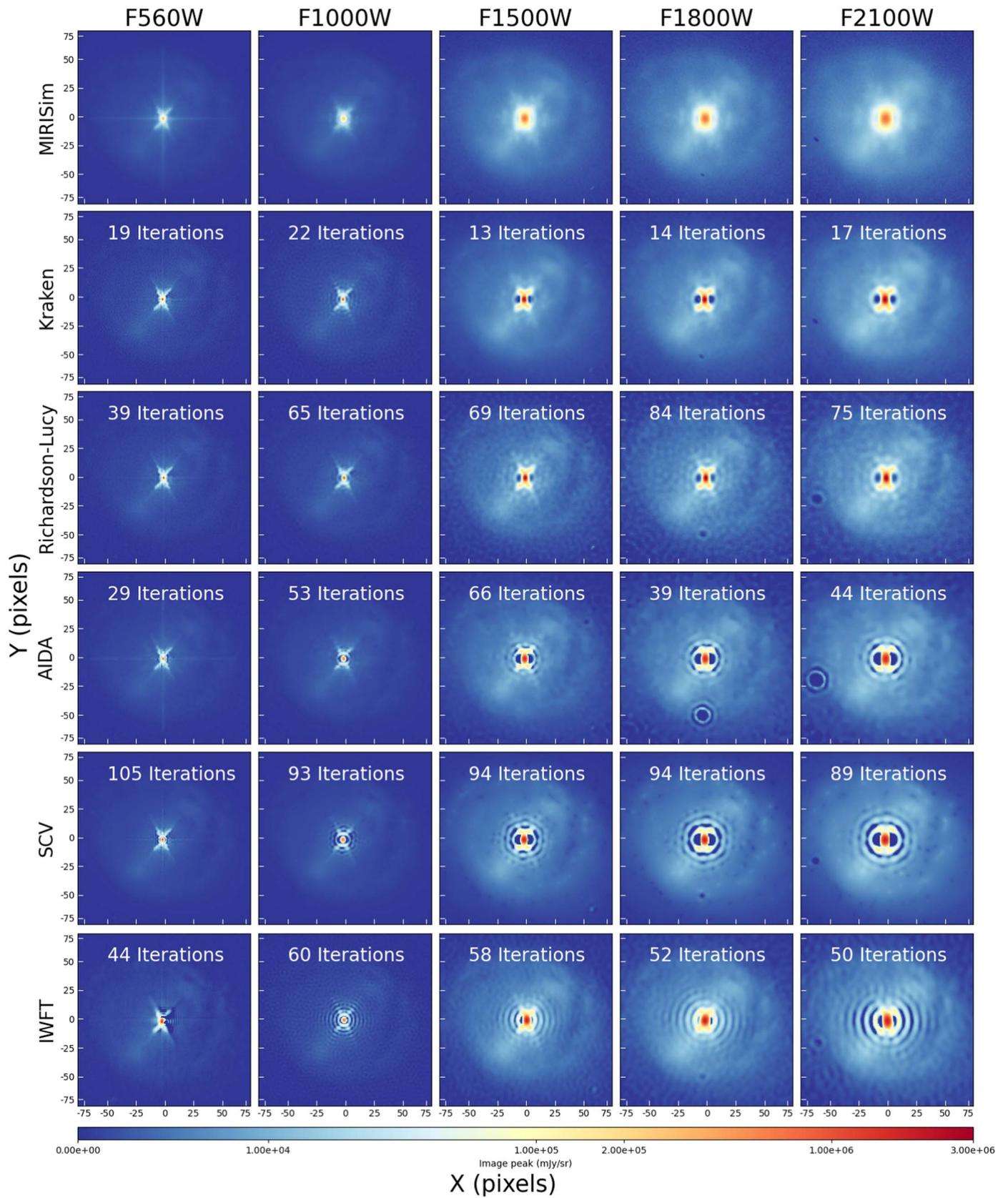

**Figure 5.** Toy AGN model image grid comparisons between the MIRISim and deconvolved images as a function of filter. Each image shows the central 150 × 150 MIRIM pixels in the *x*- and *y*-coordinate plane, is log scaled, and photometrically scaled to the IWFT F2100W image. (Top row) MIRISim, (second row) Kraken, (third row) Richardson–Lucy, (fourth row) AIDA, (fifth row) SCV, and (sixth row) IWFT. The number of deconvolution iterations the image converged to is noted in each image panel.

Table 2
Deconvolution Merit Function Results for the Simulated MIRIM Observations of the Toy AGN Model

| Filter | F560W | F1000W | F1500W | F1800W | F2100W |
|---|---|---|---|---|---|
| MIRISim | | | | | |
| FWHM (pixel) | 2.78 | 4.14 | 6.68 | 7.56 | 8.74 |
| Aperture flux (mJy sr$^{-1}$) | 8.85E+04 | 1.02E+05 | 5.58E+05 | 7.08E+05 | 9.23E+05 |
| Kraken | | | | | |
| Iterations | 19 | 22 | 13 | 14 | 17 |
| FWHM (pixel) | 1.23 | 1.70 | 3.13 | 3.42 | 4.17 |
| $\Delta_{FWHM}$ (%) | 55.87 | 59.03 | 53.13 | 54.83 | 52.23 |
| Ratio (MIRISim/Kraken) | 2.27 | 2.44 | 2.13 | 2.21 | 2.09 |
| Aperture flux (mJy sr$^{-1}$) | 1.02E+05 | 1.13E+05 | 6.26E+05 | 8.11E+05 | 1.08E+06 |
| $\Delta_{flux}$ (%) | 15.32 | 10.73 | 11.99 | 15.00 | 16.54 |
| Ratio (MIRISim/Kraken) | 0.87 | 0.90 | 0.89 | 0.87 | 0.86 |
| Richardson–Lucy | | | | | |
| Iterations | 39 | 65 | 69 | 84 | 75 |
| FWHM (pixel) | 1.60 | 2.05 | 3.21 | 3.45 | 4.21 |
| $\Delta_{FWHM}$ (%) | 42.34 | 50.40 | 51.92 | 54.36 | 51.79 |
| Ratio (MIRISim/RL) | 1.73 | 2.02 | 2.08 | 2.19 | 2.07 |
| Aperture flux (mJy sr$^{-1}$) | 1.01E+05 | 1.11E+05 | 6.17E+05 | 7.97E+05 | 1.05E+06 |
| $\Delta_{flux}$ (%) | 14.11 | 9.19 | 10.52 | 12.73 | 13.58 |
| Ratio (MIRISim/RL) | 0.88 | 0.92 | 0.90 | 0.89 | 0.88 |
| AIDA | | | | | |
| Iterations | 29 | 53 | 66 | 39 | 44 |
| FWHM (pixel) | 1.83 | 2.39 | 4.08 | 5.01 | 5.68 |
| $\Delta_{FWHM}$ (%) | 34.20 | 42.17 | 38.99 | 33.81 | 35.03 |
| Ratio (MIRISim/AIDA) | 1.52 | 1.73 | 1.64 | 1.51 | 1.54 |
| Aperture flux (mJy sr$^{-1}$) | 8.99E+04 | 1.02E+05 | 5.69E+05 | 7.36E+05 | 9.70E+05 |
| $\Delta_{flux}$ (%) | 1.57 | 0.39 | 1.99 | 4.00 | 5.03 |
| Ratio (MIRISim/AIDA) | 0.99 | 1.0 | 0.98 | 0.96 | 0.95 |
| SCV | | | | | |
| Iterations | 105 | 93 | 94 | 94 | 89 |
| FWHM (pixel) | 1.61 | 2.59 | 4.09 | 4.76 | 5.56 |
| $\Delta_{FWHM}$ (%) | 41.97 | 37.38 | 38.83 | 37.03 | 36.42 |
| Ratio (MIRISim/SCV) | 1.72 | 1.60 | 1.63 | 1.59 | 1.57 |
| Aperture flux (mJy sr$^{-1}$) | 1.02E+05 | 1.13E+05 | 6.29E+05 | 8.27E+05 | 1.07E+06 |
| $\Delta_{flux}$ (%) | 15.48 | 10.98 | 12.76 | 16.75 | 15.70 |
| Ratio (MIRISim/SCV) | 0.87 | 0.90 | 0.89 | 0.86 | 0.86 |
| IWFT | | | | | |
| Iterations | 45 | 60 | 58 | 52 | 50 |
| FWHM (pixel) | 1.64 | 2.73 | 4.32 | 5.04 | 5.95 |
| $\Delta_{FWHM}$ (%) | 41.08 | 33.89 | 35.57 | 33.38 | 31.86 |
| Ratio (MIRISim/IWFT) | 1.70 | 2.73 | 1.55 | 1.50 | 1.47 |
| Aperture flux (mJy sr$^{-1}$) | 1.02E+05 | 1.12E+05 | 6.29E+05 | 8.23E+05 | 1.07E+06 |
| $\Delta_{flux}$ (%) | 15.21 | 10.62 | 12.72 | 16.26 | 16.18 |
| Ratio (MIRISim/IWFT) | 0.87 | 0.90 | 0.89 | 0.86 | 0.86 |

**Notes.** The $\Delta$(%) and ratio values were determined between the toy AGN model MIRISim and deconvolved images for each algorithm in each filter.

Ingaramo et al. 2014). Due to the widespread usage and robustness of Richardson–Lucy we chose to include this algorithm in our comparison.

AIDA is a reimplementation and extension of the Myopic Iterative sTep Preserving ALgorithm (MISTRAL; Mugnier et al. 2004), which utilizes a Bayesian framework and iterative optimization procedure to estimate the true image and PSF (given the initial image and partially known PSF). AIDA forward models the noise (i.e., detector readout noise, shot noise, etc.) in an image and employs regularization criteria to (1) restore sharp object edges without ringing effects, (2) estimate the PSF under soft constraints (i.e., the PSF is not assumed to be precisely known but is estimated with some level of uncertainty) rather than blindly, and (3) preserve the photometry of the image. This regularization yields object reconstruction with excellent edge preservation (e.g., Storrs et al. 2005) and flux conservation, which is why we chose to include this algorithm in our comparison.

SCV is a wavelet-based deconvolution approach which utilizes a starlet transformation (Starck et al. 2015a) to capture

**Table 3**
Position Angle and Eccentricity for the Toy AGN Model MIRISim and Final Deconvolved Images in the Filters as Noted

| Filter | F560W | F1000W | F1500W | F1800W | F2100W |
|---|---|---|---|---|---|
| | | MIRISim | | | |
| PA (°) | 82.05 | 82.13 | 82.88 | 83.60 | 83.33 |
| Eccentricity | 0.47 | 0.54 | 0.51 | 0.49 | 0.46 |
| | | Kraken | | | |
| PA (°) | 82.12 | 82.55 | 82.90 | 83.41 | 84.26 |
| Eccentricity | 0.58 | 0.66 | 0.65 | 0.65 | 0.62 |
| | | Richardson–Lucy | | | |
| PA (°) | 82.43 | 82.73 | 83.35 | 83.61 | 83.86 |
| Eccentricity | 0.58 | 0.64 | 0.66 | 0.65 | 0.64 |
| | | AIDA | | | |
| PA (°) | 82.30 | 82.68 | 83.19 | 83.03 | 84.16 |
| Eccentricity | 0.54 | 0.66 | 0.65 | 0.63 | 0.63 |
| | | SCV | | | |
| PA (°) | 82.09 | 82.62 | 82.99 | 83.87 | 83.77 |
| Eccentricity | 0.58 | 0.63 | 0.66 | 0.65 | 0.66 |
| | | IWFT | | | |
| PA (°) | 82.01 | 81.81 | 82.07 | 79.89 | 81.20 |
| Eccentricity | 0.55 | 0.55 | 0.55 | 0.48 | 0.48 |

**Note.** The PA and eccentricity values for the MIRISim and final deconvolved images were determined within an aperture of 36 pixels (diameter) relative to the image array $x$-axis.

information about different spatial frequency components in the image (i.e., it decomposes the image into a series of wavelet coefficients) and sparse regularization (Starck et al. 2015b) to minimize the number of wavelet coefficients that represent the image. The deconvolution performance is improved (i.e., in terms of ringing artifact suppression, noise reduction, and source recovery) by minimizing the number of wavelet coefficients used to represent the image in the wavelet domain. The wavelets are then iteratively deconvolved using the primal-dual splitting Condat–Vũ algorithm (Vu 2011; Condat 2013). Decomposing the image into a series of wavelets allows for multiresolution analysis (i.e., analysis at different spatial frequency levels), noise reduction, and ringing artifact suppression, leading us to include this algorithm in our comparison. However, this type of deconvolution can introduce edge effects into the image and underperforms if the PSF is not well known a priori, or if the PSF is very complex.

IWFT is based on the alternating direction method of multipliers (ADMM; Boyd et al. 2011), which decomposes the ill-posed problem of deconvolution into two subproblems, solving separately for the initial image data fidelity and regularization criteria. IWFT uses two sets of filters, one for the initial image restoration (restore filter) and another for ringing artifact suppression (update filter), which is used iteratively to restore the image using TV regularization (Rudin et al. 1992). Both filters are computed for a user-defined degradation (e.g., blur and noise level) and filter size. IWFT performs well at removing ringing artifacts from synthetic imaging after only a few iterations (Šroubek et al. 2019), leading us to include this algorithm in our comparison; however, we are unaware of this algorithm's previous use in astronomy.

### 3.2. Merit Functions

To avoid introducing ringing and other artifacts into our deconvolved images, we aimed to minimize the number of iterations $n$ used for each deconvolution algorithm. To accomplish this, the following merit functions were used to serve as the regularization convergence criteria for our deconvolution methodology.[46] When either of them is reached the deconvolution iteration is terminated and the final image is returned.

1. $\Delta FWHM$: the ΔFWHM of the nucleus between consecutive iterations converges to <0.1%.
2. *Flux conservation*: the Δflux in a given aperture between the original and the $n$th-deconvolved image diverges by >30%.

The centroid of the source was first determined in a 45 pixel (5″) box around the nucleus using a quadratic centroiding algorithm. For the first merit function, we found measuring the FWHM by fitting a 1D Guassian in a small aperture (12 pixels, 1″.4) at that centroid, in both $x$ and $y$ in the frame coordinates of the array, and then averaged to be the most robust measurement. The small aperture was used to avoid contamination from the ionization bicone component. With increasing deconvolution iterations, the initially large ΔFWHM reduced as the algorithm converged to a solution.

For the second merit function an aperture of sufficient size to enclose the dominant flux from the PSF but with minimal noise and other source contamination was used for all wavelengths. The aperture was sized for the F2100W filter and contained ∼75% of the flux of the PSF, with the remaining ∼25% in the extended complex PSF beyond the first Airy ring (Rigby et al. 2023). An aperture size that enclosed the first Airy ring at the longest wavelength was chosen to ensure a valid comparison at all wavelengths (diameter of 36 pixels, 4″.1). If we had used a smaller aperture (i.e., to enclose only the Airy disk), as the image quality improved with deconvolution iteration, the flux from the first Airy ring could be included in that aperture, thus significantly increasing the flux. For each iteration of the deconvolution and in each filter, the aperture was positioned at the centroid of the source and the flux measured. The merit function for flux conservation was set to 30% as the iteration terminus as ∼25% of the flux was outside the first Airy ring and we afforded an additional 5% as a buffer.

### 4. Deconvolved Toy AGN Model Results

In this section we present a comparison of the key results for the MIRIM simulations of our toy AGN model and then the results for the five employed deconvolution algorithms. As a key component of our work is the comparison of each algorithm's performance across the five simulated wavelength images we chose to forgo the simultaneous deconvolution of all five images and instead deconvolved each image to convergence based on our merit functions (Section 3.2).

Figure 5 shows the input MIRISim toy AGN model and deconvolved images for each deconvolution algorithm and for each of the five filters after reaching the regularization convergence criteria. The MIRISim simulated observations (top row) clearly show the ionization bicone in the F560W and F1000W filters, but a "boxy" morphology centered on the Airy disk is dominant in the other three filters. In all but the shortest-wavelength filter, the complex JWST PSF can be clearly observed projected onto the host galaxy. Table 2 shows the merit function measurements, Δ(%), and ratio values (between the MIRISim and deconvolved merit functions) for the input MIRISim toy AGN model and deconvolved images in each filter. The FWHM increases essentially in synergy with the central wavelength of the filter for the MIRISim imaging (Table 2) as expected from the physics of diffraction. Table 3 shows the PAs and eccentricity values for the input MIRISim toy AGN model and deconvolved images in each filter, determined using the Photutils ApertureStats[47] package. In all cases for the MIRISim imaging the polar dust component is not observed, but the presence of the component is indicated by the PA and eccentricity (Table 3, over and above the eccentricity from the ionization bicone as measured in the same sized aperture used for photometry) and as expected the eccentricity reduces with longer wavelength (except for the F560W filter), consistent with the increasing diffraction limit reducing the model's spatial resolution. The F560W image exhibits the crosshair effect expected for this filter (Section 2.1, Figure 1), likely skewing the eccentricity measurement for this image. The host galaxy is most obviously seen in the longest wavelengths due to the image scaling as the image peak sets the scaling at that wavelength for all algorithms (the peak is most concentrated in the shorter wavelengths due to the diffraction limit of the telescope).

---
[46] https://github.com/MTLeist255/JWST_Deconvolution

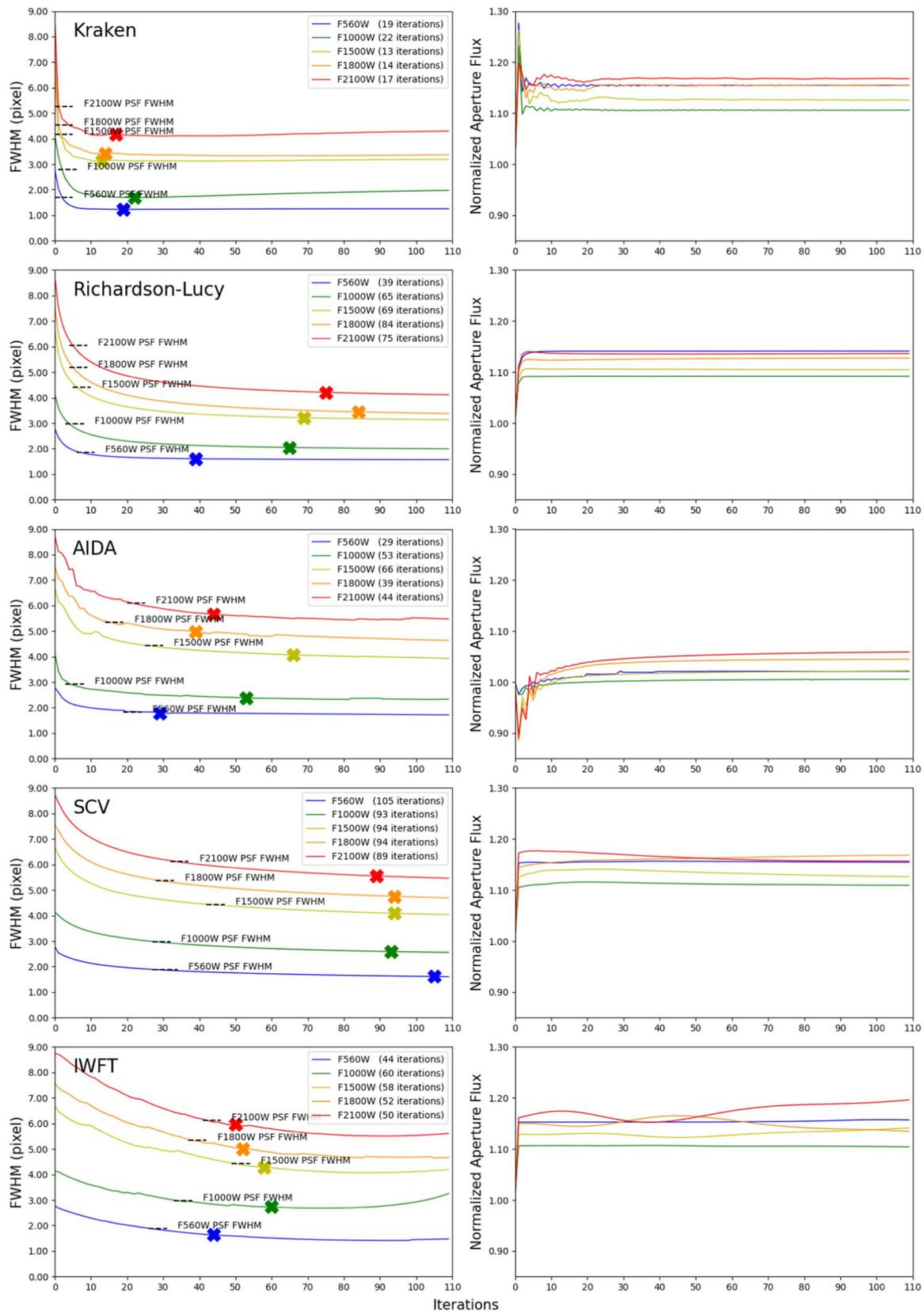

**Figure 6.** Merit function results for simulated MIRIM observations of the toy AGN model. Left: FWHM results as a function of deconvolution iteration. The iteration number where the measured FWHM value dropped below the theoretical diffraction limit for that image is marked with a dotted line. The iteration number where the regularization convergence criteria were met is listed in the legend and marked by an "X" for each plot. Right: normalized flux in the aperture as a function of deconvolution iteration. The aperture flux for each deconvolved image was normalized to the MIRISim image in the same wave band. Note the increase in normalized aperture flux after the first few iterations for each algorithm caused by the initial concentration of flux within the aperture.

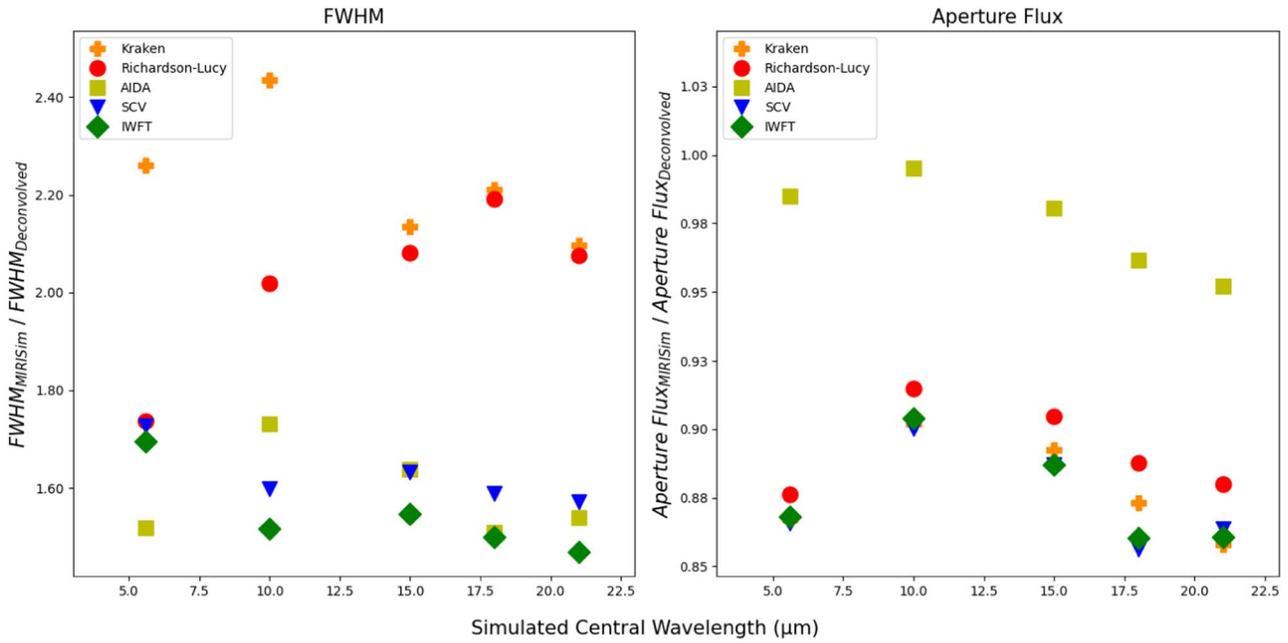

**Figure 7.** Ratio between the MIRISim and final deconvolved toy AGN model images for each deconvolution algorithm as a function of simulated central wavelength. Each point is plotted at the central wavelength for the respective filter. Left: FWHM ratios. Right: aperture flux ratios.

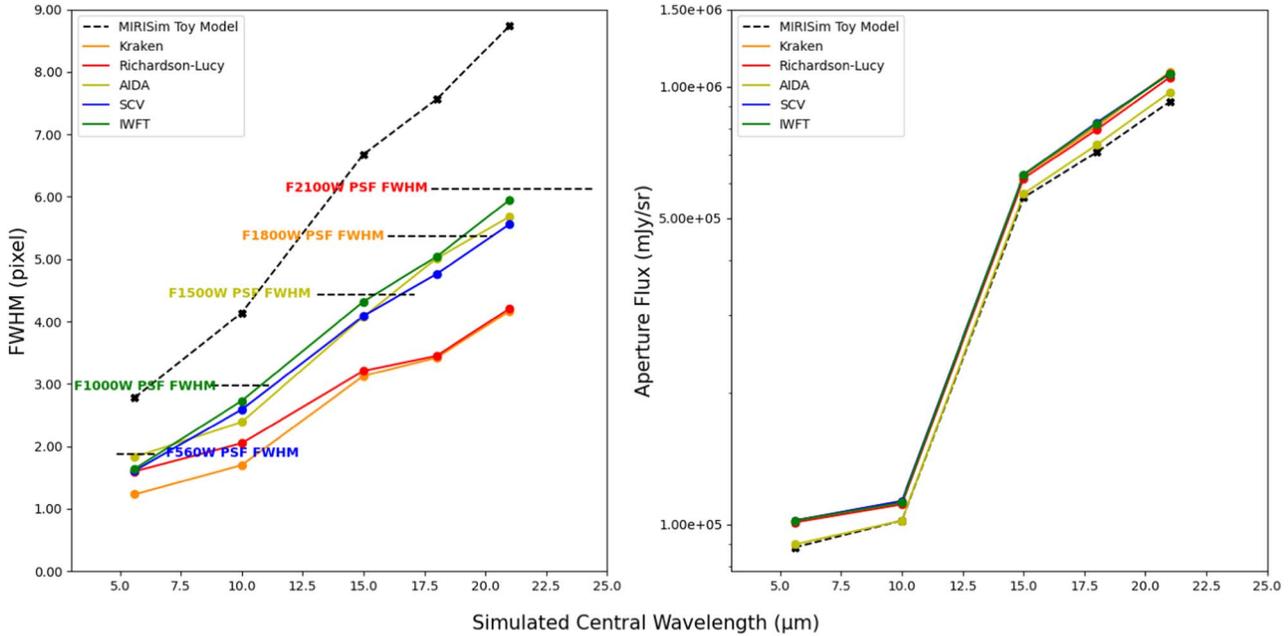

**Figure 8.** FWHM and aperture flux comparisons between MIRISim and the deconvolved images as a function of simulated central wavelength. Each point is plotted at the central wavelength for the respective filter. Left: measured FWHMs, where the dotted line shows the theoretical PSF FWHM at the center of the respective filter, and the length of the dotted line represents the filter width. Right: log-scaled comparative SEDs generated for the MIRISim and deconvolved images.

---

[47] https://photutils.readthedocs.io/en/stable/api/photutils.aperture.ApertureStats.html

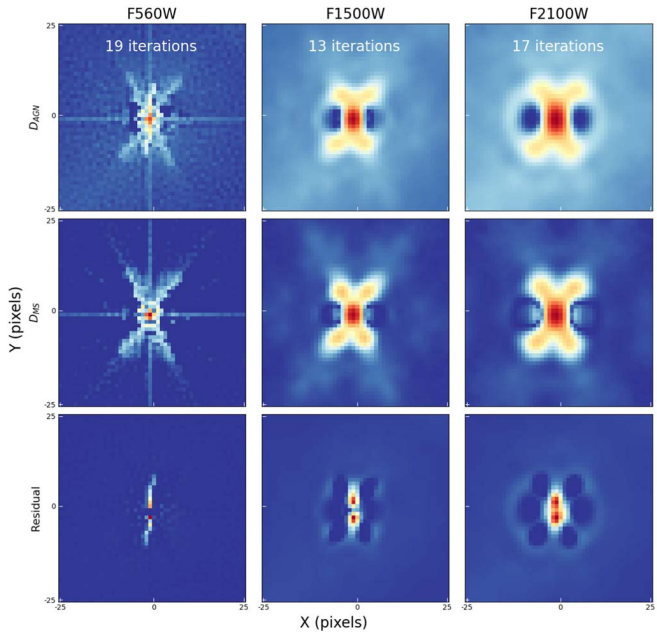

**Figure 9.** Kraken deconvolved toy AGN $D_{AGN}$ (top row), Kraken deconvolved central region combined with ionization bicone $D_{MS}$ (middle row), and residual images (bottom row). The top two rows are displayed log scaled and the bottom row is displayed in linear scale. Each image is photometrically scaled to the F2100W image in each row and shows the central $50 \times 50$ MIRIM pixels.

The Kraken deconvolved images (second row, Figure 5) show little evidence of ringing or orange peel effects (except in the F560W and F1000W filters). The first Airy ring is well removed, and the ionization bicone is recovered across all filters, but the polar dust component remains visually undetected except in the F560W image. In this image a partial visual detection of an elongation similar in morphology to the polar dust component is detected aligned with the PA of the residual crosshair effect observed at this wavelength. Figure 6 shows the merit function results as a function of deconvolution iteration for each algorithm. The central region's FWHM is reduced by as much as a factor of 2.4, but $\Delta$flux shows an increase of up to 16%. The eccentricity due to the ionization bicone and polar dust is higher than the simulated MIRIM image across all five filters, but maintains a similar PA. It is possible to reduce the ringing and orange peel effects observed in the shortest-wavelength filter images further at the marginal expense of FWHM. However, as the merit function was optimized for the FWHM, we accept the additional orange peel effect for the purposes of this work.

Richardson–Lucy deconvolution showed a good overall result with a relatively high number of iterations as compared to Kraken. At all wavelengths the first Airy ring is well removed, the ionization bicone is well recovered, and again find no clear recovery of the polar dust component (except in the F560W image, which could also be a residual of the crosshair effect; third row, Figure 5). The FWHM of the central source is reduced by up to a factor of 2.2 and $\Delta$flux increases by as much as 14%. Especially in the longer-wavelength filters there is clear evidence of both ringing and orange peel effects, a sign of overdeconvolution near the central region in the plane perpendicular to the ionization bicone. The eccentricity due to the ionization bicone and polar dust is higher than the simulated MIRIM image across all five filters, but maintains a similar PA.

AIDA suffers from significant ringing effects near the nucleus, which are visually clearer in the longest-wavelength filters due to the image scaling (fourth row, Figure 5). Around the central region there is some evidence of Airy rings remaining present, which combined with the ringing make recovery of the ionization bicone challenging in the longest-wavelength filters but is clearly observed in the shorter filters. The FWHM reduces by a factor of 1.7 but $\Delta$flux only increases by as little as 0.4%.

SCV converged after the highest number of iterations and is also visually dominated by strong ringing effects near the central region, which are visually clearer in the longest-wavelength filters but present in all filters (fifth row, Figure 5). The ringing effects serve to render the ionization bicone difficult to detect in the longer-wavelength filters. The FWHM of the central source is improved by a factor of up to 1.7 and $\Delta$flux is improved by as much as 17%.

Finally, the IWFT algorithm is visually dominated by strong ringing effects near the central region, where neither the first Airy ring nor ionization bicone are well recovered. Indeed in the longest-wavelength filters multiple Airy rings remain (sixth row, Figure 5). The FWHM of the central source is improved by a factor of 1.7, and $\Delta$flux is improved by as much as 16%.

The FWHM and flux results for all five methodologies are compared in Figure 7. At all wavelengths Kraken shows the best improvement in the FWHM, whereas the optimal flux conservation is that of AIDA. Based on the visual representation of the images and FWHM improvements we argue that Kraken is the optimal deconvolution algorithm followed by Richardson–Lucy. We note that Kraken and Richardson–Lucy afford similar FWHM improvement at longer wavelengths, but Kraken provides a significantly better result for shorter-wavelength filters.

Figure 8 shows the combined FWHM and flux for all algorithms and filters. The FWHM versus wavelength plot also shows that the Richardson–Lucy and Kraken algorithms have the greatest FWHM improvement, and again demonstrates the superiority of Kraken at the shortest wavelengths. The flux versus wavelength plot shows both the adopted SED (Section 2.2) as well as the flux increase from the MIRIM simulations to the deconvolved images, showing an increasing flux in synergy with increasing wavelength, except for the F560W filter. For the F560W image the photometric aperture includes more emission from the host galaxy, which in this filter is bright and complex compared to the longer-wavelength filters. That the flux increase is correlated with increasing wavelength is indicative of continued concentration of the flux within the photometric aperture rather than an additional deconvolution artifact. To probe this further, we simulated a simple PSF using MIRISim and then deconvolved using Richardson–Lucy to a similar number of iterations used for our deconvolution comparison. Using a large aperture (diameter 11″) to include the dominant amount of flux from the PSF (Figure 1), we measured the flux as a function of iteration and found an increase of only <0.45% from the original simulated image in all five filters, showing that Richardson–Lucy does not artificially increase the flux. Instead, we suggest that the increased flux measured as a function of deconvolution iteration is a result of additional extended flux being included within the photometric aperture.

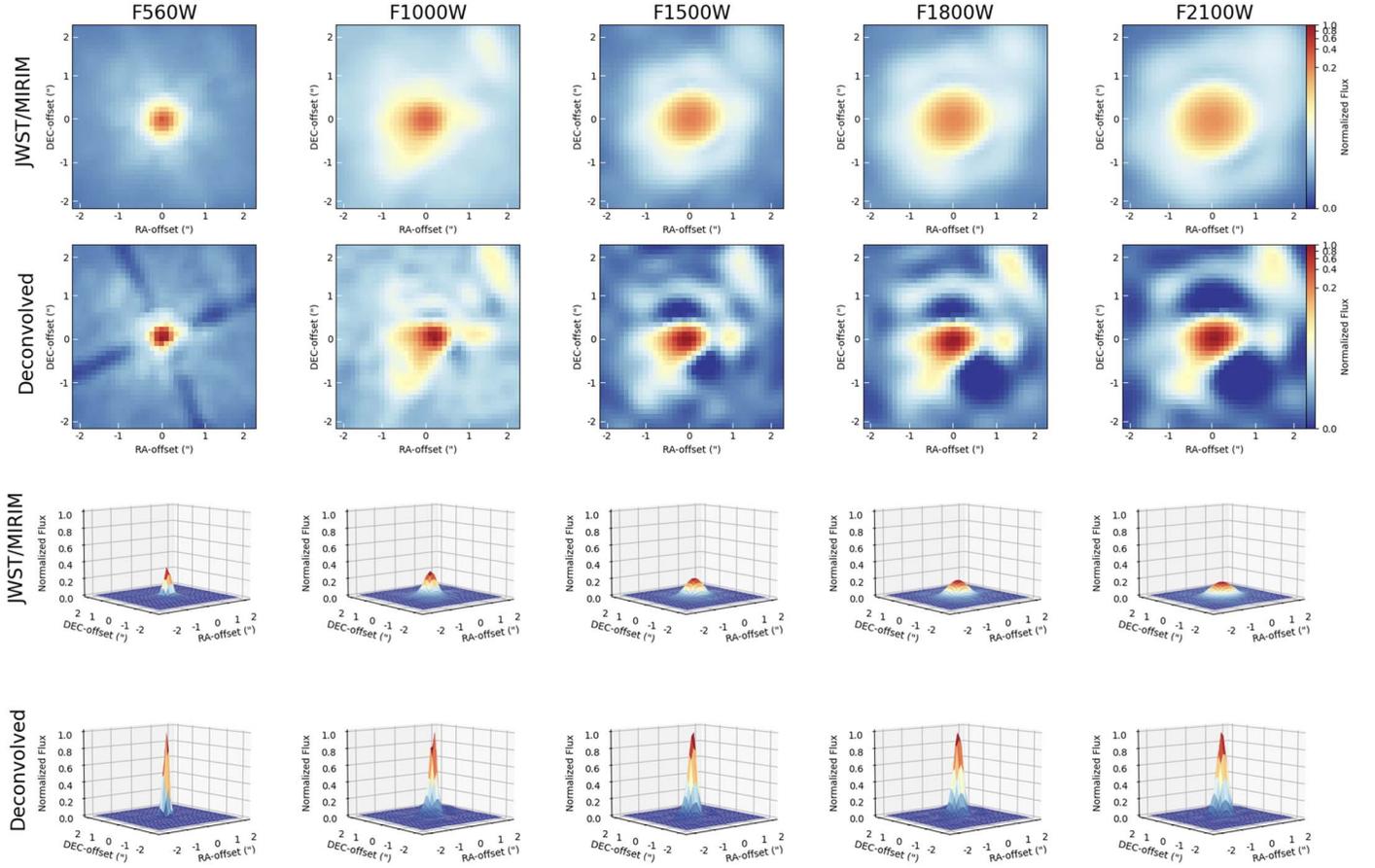

**Figure 10.** JWST/MIRIM (row 1) and Kraken deconvolved (row 2) images for each filter. Each image is normalized, displayed log scaled, scaled to the corresponding deconvolved image peak for each filter, rotated to the N through E orientation, and ∼4″ × 4″. Rows 3 and 4 display the normalized flux distribution of the central ∼4″ × 4″ for the JWST/MIRIM and Kraken deconvolved images, respectively.

**Table 4**
Kraken Deconvolution Results for NGC 5728

| Filter | F560W | F1000W | F1500W | F1800W | F2100W |
| --- | --- | --- | --- | --- | --- |
| | | JWST/MIRIM | | | |
| FWHM (″) | 0.30 | 0.56 | 0.63 | 0.75 | 0.83 |
| Aperture flux (mJy sr$^{-1}$) | 8.84E+04 | 1.02E+05 | 5.57E+05 | 7.05E+05 | 9.02E+05 |
| | | Kraken | | | |
| Iterations | 6 | 18 | 19 | 20 | 18 |
| FWHM (″) | 0.18 | 0.27 | 0.29 | 0.34 | 0.37 |
| $\Delta_{\mathrm{FWHM}}$ (%) | 40.00 | 52.54 | 53.30 | 54.25 | 55.26 |
| Ratio (Observed/deconvolved) | 1.60 | 2.11 | 2.14 | 2.19 | 2.24 |
| Aperture flux (mJy sr$^{-1}$) | 1.02E+05 | 1.12E+05 | 6.36E+05 | 8.20E+05 | 1.08E+06 |
| $\Delta_{\mathrm{flux}}$ (%) | 14.96 | 10.13 | 14.07 | 16.32 | 17.04 |
| Ratio (Observed/deconvolved) | 0.87 | 0.91 | 0.88 | 0.86 | 0.85 |

Despite the image quality being improved by deconvolution, there was no clear visual detection of the polar dust component in any of the images. To examine further if deconvolution can assist in the detection of polar dust, we generated a model consisting of only the brightest central components (i.e., the central region PSF and ionization bicone), maintaining the same physical extent and intensity as used in the toy AGN model (Section 2.2). We input this into MIRISim in the highest, mid, and lowest spatial resolution filters (F560W, F1500W, and F2100W, respectively), generated MIRIM simulated images (Section 2.2), Kraken deconvolved these MIRISim images ($D_{\mathrm{MS}}$) to the same number of iterations as the Kraken deconvolved toy AGN model ($D_{\mathrm{AGN}}$), then subtracted $D_{\mathrm{MS}}$

Table 5
Position Angle and Eccentricity for the JWST/MIRIM and Kraken Deconvolved Images of NGC 5728

| Filter | F560W | F1000W | F1500W | F1800W | F2100W |
|---|---|---|---|---|---|
| | | JWST/MIRIM | | | |
| PA (°) | 115.49 | 123.24 | 110.90 | 111.26 | 109.69 |
| Eccentricity | 0.57 | 0.51 | 0.62 | 0.62 | 0.66 |
| | | Kraken | | | |
| PA (°) | 115.66 | 124.56 | 111.22 | 112.59 | 111.0 |
| Eccentricity | 0.61 | 0.53 | 0.68 | 0.69 | 0.77 |

**Notes.** The PA and eccentricity values were determined in a 4″.1 aperture, measured N through E.

from $D_{AGN}$ to give residual model images for each filter. Figure 9 shows $D_{AGN}$, $D_{MS}$, and the residual images, where the residual images show an elongation consistent with the modeled polar dust component in all three filter images.

All the deconvolution algorithms we used improved, some significantly, image quality (i.e., reduced the FWHM below the theoretical JWST/MIRIM diffraction limit for each filter image, presence of the first Airy ring, etc.). Kraken afforded the greatest FWHM reduction with good photometric integrity across all five simulated filters and introduced the fewest artifacts. For these reasons, we selected this algorithm to apply to our JWST/MIRIM observations, as described in Section 5. Additional deconvolution tests assessing the validity of our results are discussed in Appendix B.

## 5. Application to JWST/MIRIM Observations

In this section we present comparisons of the key results for Kraken deconvolution of JWST/MIRIM observations of NGC 5728. Detector-sampled PSF models generated in Section 2.1 (Figure 1) were used as the reference PSFs for deconvolution. The top two rows of Figure 10 show the five filter JWST/MIRIM and Kraken deconvolved (after reaching the regularization convergence criteria; Section 3.2) central ∼4″ × 4″ images, respectively. The lower two rows of Figure 10 show the 3D normalized flux distributions within each of the central ∼4″ × 4″ of the JWST/MIRIM and Kraken deconvolved images, respectively. The significant improvement in the central source, the first Airy ring reduction, and extended galaxy is readily observed from a visual examination.

We visually detect a clear ∼2″.5 (∼470 pc) extended nuclear emission, extending SE to NW centered on the nucleus with an average central PA (across all five bands) ∼115°, which was not readily visible in the JWST/MIRIM images prior to deconvolution, especially in the longest-wavelength filters. However, the crosshair pattern in the F560W filter is notably worse in the deconvolved image, possibly due to the bright central region not being a true point source and the image being undersampled at this wavelength, thereby complicating the Kraken deconvolution. We note the strong correspondence of features and morphology in the images (Figure 10) and the good morphological correspondence to the 130 pc SE and 230 pc NW AGN-driven weak outflow observed for this object (Shimizu et al. 2019), further increasing confidence in the validity of the converged deconvolved images. A full interpretation of the morphology, SED, and other astrophysical results is beyond the scope of this paper and is instead covered in D. J. V. Rosario et al. (2024, in preparation).

Table 4 shows the merit function measurements, Δ(%), and ratio for the JWST/MIRIM and Kraken deconvolved images in each filter, and Table 5 shows the PAs and eccentricity values. Kraken reduces the FWHM of the nuclear source by a factor of 1.6–2.2 across all five filters used. The FWHM improvement of up to a factor of 2.2 is lower than the toy model AGN improvement, indicative of the central source not being only a point source. This is consistent with the residuals of the first Airy ring being imperfectly deconvolved, most clearly ∼1″ N of the nucleus (in the longest-wavelength filter images, Figure 10), and with a greater residual than our point-source-dominated toy AGN model.

Figure 11 shows the merit function results as a function of deconvolution iteration for the Kraken deconvolved images. The FWHM versus iteration plot shows the significant improvement of the central source FWHM, which is below the theoretical JWST/MIRIM diffraction limit for each filter image. We find a 10%–20% increase in flux in an aperture of 4″.1, which is consistent with the results of the toy model and thus we consider the flux increase for NGC 5728 as consistent with extended flux being deconvolved to within the photometric aperture. Figure 12 shows the FWHM and flux measurements through each filter, demonstrating the FWHM improvement and good flux conservation. Figure 13 shows a photometrically accurate five color image (based on the aperture flux in Table 4) for the JWST/MIRIM and Kraken deconvolved images. The crosshair pattern most readily detected in the F560W filter is clearly observed in both figures. In the deconvolved image the first Airy ring is reduced and in the innermost region some residual color effects remain. Finally, the ∼2″.5 SE to NW nuclear extension is clearly observed in the deconvolved image, demonstrating the success of Kraken at enhancing this feature.

## 6. Summary and Conclusions

In this paper, we described a test methodology and results of deconvolution algorithms on simulated and observed JWST/MIRIM images of AGNs in five filters. We concentrated on the simulataneous image quality improvement of a bright point source surrounded by low surface brightness emission. We generated MIRI SUB256 sampled PSFs utilizing WebbPSF for the F560W, F1000W, F1500W, F1800W, and F2100W filters, using these as reference PSFs for model and observation deconvolution. These model PSFs included the crosshair effect but did not include effects from the MIRI detector, such as read noise, bad pixels, cosmic rays, etc.

We generated a toy AGN model consisting of four components: (1) a <20 pc central region (dominantly the obscuring torus illuminated by the accretion disk of the central engine) modeled as a WebbPSF MIRI detector-sampled PSF for the filters above (central region PSF) (2) an ∼200 pc elongated dusty polar extension, (3) an extended dusty ionization bicone extending hundreds of parsecs from the central region, and (4) an ∼5.3 kpc scale host galaxy. The polar dust, bicone, and galaxy components were first combined into a single, three-component model; then, using MIRISim, we performed a set of MIRI imaging simulations of the three-component model utilizing the MIRI SUB256 for the filters above. The filter-dependent central region PSF was then added to each MIRISim output, and the combined image was flux calibrated to the aperture photometric measurements of NGC 5728 in the appropriate filter. Finally, flux-calibrated MIRIM simulated read noise was added and negative pixels were masked, creating the final high-S/N images of the toy AGN model in the five filters.

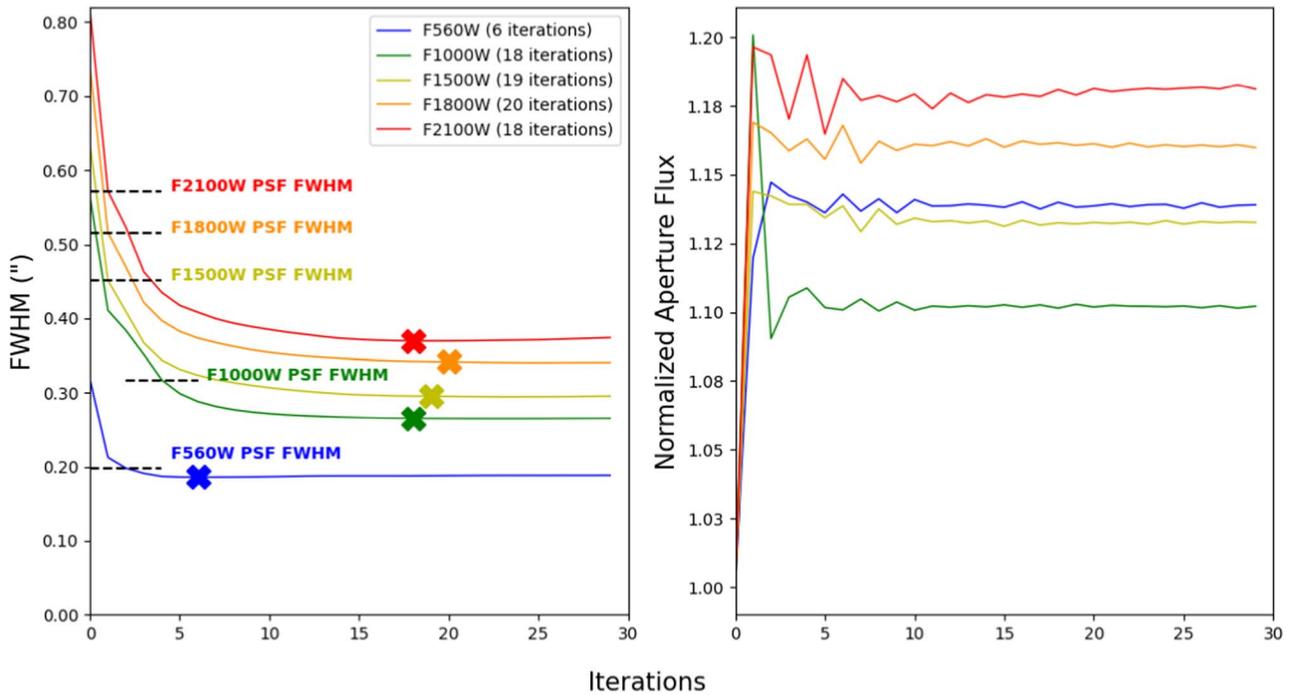

**Figure 11.** Same as Figure 6 for the Kraken deconvolved images of NGC 5728, except the FWHM is expressed in arcseconds. Aperture flux measurements were normalized to the JWST/MIRIM image in the same wave band.

Each MIRIM toy AGN model simulated image was deconvolved using five iterative deconvolution algorithms: (1) Kraken deconvolution, (2) linear regularization using noncirculant Richardson–Lucy, (3) myopic deconvolution using AIDA, (4) sparse regularization using the SCV algorithm, and (5) Wiener filtering using IWFT. The regularization convergence criteria were (1) the ΔFWHM of the nucleus between consecutive iterations converges to <0.1% and (2) the Δflux in a given aperture between the simulated and $n$th-deconvolved image diverges to >30%. Our results are summarized below.

1. In the MIRISim images (top row, Figure 5) the ionization bicone is clearly visible in the F560W and F1000W filters. The F560W image additionally shows the simulated crosshair effect expected for this filter, and the JWST PSF can be clearly observed in all but the F560W image. In all cases the polar dust is not visually detected, but the PA and eccentricity are consistent with the existence of an ∼200 pc scale polar elongation (Table 3).
2. Each algorithm improved the image quality (i.e., FWHM and first Airy ring reduction), however none were able to visually recover the polar dust component fully in any of the five simulated wave bands, although the presence of this component is indicated in the PA and eccentricity (Table 3). Kraken showed the best improvement in FWHM at all wave bands (up to a factor of 2.4), showed little evidence of ringing or orange peel effects except in the F1000W image, effectively removed the first Airy ring, recovered the ionization bicone, and maintained good photometric integrity. For these reasons, Kraken was selected to be used for our JWST/MIRIM observations.
3. Multifilter JWST/MIRI imaging of NGC 5728 were used for the Kraken deconvolution application. Kraken improved the central source, reduced the first Airy ring, and visually improved the extended galaxy in each image. The flux shows a 10%–20% increase (consistent with the Kraken toy AGN model deconvolved results; Table 5), but the FWHM improvement of up to a factor of 2.2 is lower than obtained with the model, consistent with the central source not being a point source. This is supported by the F560W filter crosshair pattern appearing visually worse in the deconvolved image, the residuals of the first Airy ring being imperfectly deconvolved, and the presence of an ∼2″.5 SE–NW nuclear extension (Figure 13). However, we note that (1) the central source FWHM converged below the diffraction limit in all five filters and (2) there is a strong correspondence of features and morphology in all five images (Figure 10), giving confidence to the validity of the converged deconvolution.

In conclusion, using deconvolution algorithms, characterizing the PSF, reducing the observational data through the JWST pipeline, and defining a set of regularization convergence criteria, we showed that deconvolution techniques can be effective for JWST/MIRI imaging to produce improved imaging results. Kraken was the best deconvolution technique we tested based on our metrics. We are optimistic that similar deconvolution can be applied to broader science cases using JWST/MIRI imaging observations. We have not yet tested deconvolution at shorter wavelengths using different JWST instruments (e.g., the Near-Infrared Camera/Near InfraRed Imager and Slitless Spectrograph) nor other MIRI observing modes (e.g., MIRI Medium Resolution Spectroscopy).

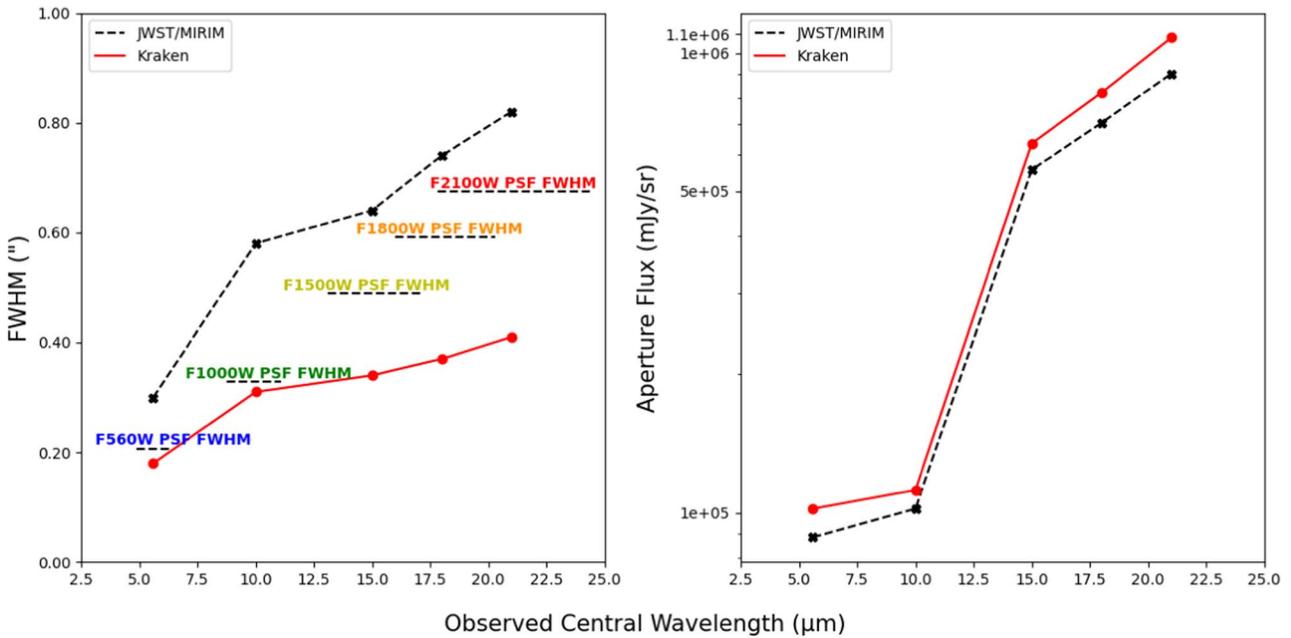

**Figure 12.** Left: FWHM of the nuclear source of NGC 5728 as a function of observed central wavelength. The dotted line shows the JWST/MIRIM PSF FWHM at the center of the respective filter, and the length of the dotted line represents the filter width. Right: comparative log-scaled aperture flux measurements of the central $4\rlap.{''}1$ between the JWST/MIRIM and Kraken deconvolved images.

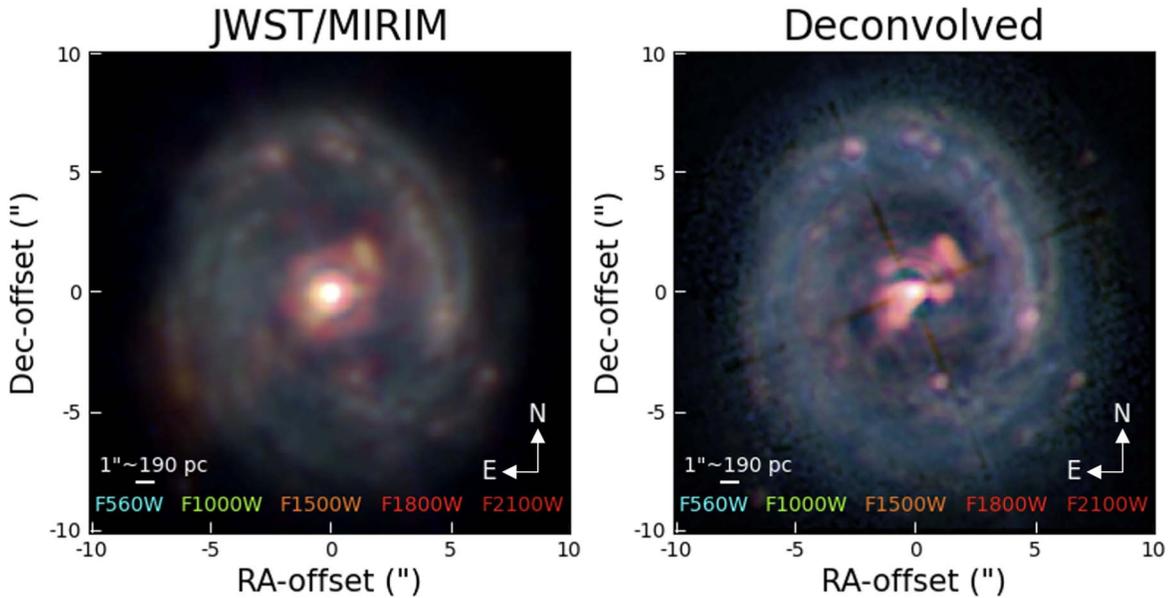

**Figure 13.** Five color JWST/MIRIM (left) and Kraken deconvolved images (right). Each image is ∼20″ × 20″, oriented N–E, and log scaled. Note the ∼$2\rlap.{''}5$ SE to NW extended nuclear emission visible in the deconvolved image. Evidence of the Airy ring remnant (∼1″ NW of the nucleus) and the F560W crosshair are also clearly visible in the deconvolved image.

However, at shorter wavelengths the PSF is more compact and hence deconvolution seems likely to be less effective.

### Acknowledgments

We thank Enrique López-Rodríguez, and the anonymous referee for providing helpful input and comments that improved the manuscript. We also thank Julian Christou for valuable insight at the intitation of this work. This work is based [in part] on observations made with the NASA/ESA/CSA James Webb Space Telescope data. The JWST data were obtained from the Mikulski Archive for Space Telescopes at the Space Telescope Science Institute, which is operated by the Association of Universities for Research in Astronomy, Inc., under NASA contract NAS 5-03127 for JWST. The JWST observations are associated with program #2064, the specific observations analyzed can be accessed via doi:10.17909/d1my-ya64. This work is based [in part] on observations made with the NASA/ESA Hubble Space Telescope, and obtained from the Hubble Legacy Archive, which is a collaboration between the Space Telescope Science Institute (STScI/NASA), the Space Telescope European Coordinating Facility (ST-ECF/ESAC/ESA) and the Canadian Astronomy Data Centre (CADC/NRC/CSA). The specific HST observation used for the toy AGN model can be accessed via

doi:10.17909/yjb7-aa73. M.T.L., C.P., E.H.S., and L.Z. acknowledge grant support from the Space Telescope Science Institute (ID: JWST-GO-02064.002), M.T.L. also acknowledges support from the Department of Physics & Astronomy at UTSA. A.A.H. acknowledges financial support from grant PID2021-124665NB-I00 funded by MCIN/AEI/10.13039/501100011033 and by ERDF "A way of making Europe." E.B. acknowledges the María Zambrano program of the Spanish Ministerio de Universidades funded by the Next Generation European Union and is also partly supported by grant RTI2018-096188-B-I00 funded by the Spanish Ministry of Science and Innovation/State Agency of Research MCIN/AEI/10.13039/501100011033. S.G.B. acknowledges support from the research project PID2019-106027GA-C44 of the Spanish Ministerio de Ciencia e Innovación. B.G.-L. acknowledges support from grants PID2019-107010GB-100 and PID2022-140483NB-C21, and the Severo Ochoa CEX2019-000920-S funded by MICINN-AEI/10.13039/501100011033. S. M.J. acknowledges support from AFOSR Award FA9550-14-1-0178. M.P.S. acknowledges funding support from the Ramón y Cajal program of the Spanish Ministerio de Ciencia e Innovación (RY2021-033094-I). C.R.A. acknowledges the projects "Feeding and feedback in active galaxies," with reference PID2019-106027GB-C42, funded by MICINN-AEI/10.13039/501100011033 and "Quantifying the impact of quasar feedback on galaxy evolution," with reference EUR2020-112266, funded by MICINN-AEI/10.13039/501100011033 and the European Union NextGenerationEU/PRTR. C.R. acknowledges support from Fondecyt Regular grant 1230345 and ANID BASAL project FB210003. M.S. acknowledges the support of the Ministry of Science, Technological Development, and Innovations of the Republic of Serbia through the contract No. 451-03-9/2023-14/200002. M.T.L. acknowledges the use of the following open-source software.

*Software:* Astropy (Astropy Collaboration et al. 2018), Jdaviz (Developers et al. 2023), JWST Calibration Pipeline (Bushouse et al. 2023), MIRISim (Klaassen et al. 2021), PHOTUTILS (Bradley et al. 2022), and WebbPSF (Perrin et al. 2012).

## Appendix A
## Deconvolution Parameters

For each algorithm tested we set the maximum number of iterations to 250. The inputs for Kraken include the input image, PSF, and a data mask for weighting Fourier components in the object estimate based on the diffraction cutoff at each observation wavelength. The object estimate is convolved at each iteration with the corresponding PSF to produce the data model. Kraken uses a nonlinear conjugate gradient method to minimize a deconvolution metric based on the squared difference between the image and data model computed using pixels with sufficient S/N. Merit function metrics (Section 3.1) were calculated from the outputs at each iteration to obtain a stopping criterion and thus enforce a regularization by truncation of iteration number. Noncircular Richardson–Lucy had the fewest number of user-definable inputs, taking only the input image, PSF image, and number of iterations as inputs. We utilized AIDA in its "classical" mode (i.e., PSF known a priori) and deconvolved each filter image using the default settings. Again, with assistance from the code's author, we modified the source code to output each deconvolution iteration to a FITS file for merit function testing purposes. Like noncircular Richardson–Lucy, SCV also accepts few user-defined inputs (input image, PSF image, number of iterations, and reweighting parameter). The reweighting parameter is used to counteract the bias introduced by soft thresholding (i.e., denoising) in sparse regularization (Farrens et al. 2020). We found the best balance between PSF reduction and ringing artifact suppression by setting this parameter to null for each filter. For IWFT, restore and update filters of different sizes were used, and the noise level parameter ($\gamma$) varied, until a solution with the smallest trade-off between ringing artifact suppression, FWHM reduction, and flux conservation was met for each simulated image. Each deconvolution iteration was again output to a FITS file, and merit functions measured. . The average runtime per iteration for each algorithm using a macOS Monterey (version 12.6.7) 3.7 GHz 6-core Intel Core i5 processor with 16 GB 2667 MHz DDR4 memory was (1) 5.51 seconds per iteration (s iter$^{-1}$) for Kraken, (2) 0.47 s iter$^{-1}$ for Richardson–Lucy, (3) 0.37 s iter$^{-1}$ for AIDA, (4) 10.52 s iter$^{-1}$ for SCV, and (5) 0.21 s iter$^{-1}$ for IWFT.

## Appendix B
## Additional Deconvolution Tests

### B.1. Different Model Component Configurations

To explore further the validity of the deconvolution results (Section 4) we generated two variations of the toy AGN model (hereafter the standard model; Section 2.2). Each model variation maintained the same model parameters used in the standard model (Table 1) but we sequentially changed only one parameter of the ionization bicone component. In the first model we changed the opening angle of the ionization bicone to 90° ($V_1$), and in the second model we changed the total integrated counts ratio to 100:1 compared to the central region PSF ($V_2$). Five filter MIRIM simulated (MS hereafter) images for $V_{1-2}$ were generated (MS$_{V1-2}$; Section 2.2) then Kraken ($K_{V1-2}$) and Richardson–Lucy (RL$_{V1-2}$) deconvolved to the same number of iterations as for the deconvolved standard model (Table 3) respectively for each filter. Kraken and Richardson–Lucy were selected for these additional tests as these algorithms gave the best deconvolution results of the five algorithms tested on our standard model (in terms of FWHM reduction and flux conservation) while returning the highest image quality (i.e., first Airy ring reduction, ionization bicone recovery, and limited ringing/orange peel effects). We give a comparison of the key results for MS$_{V1-2}$, $K_{V1-2}$, and RL$_{V1-2}$ below.

The MS$_{V1}$ observations (Figure B1) clearly show the ionization bicone in the F560W and F1000W filters, but a "boxy" morphology centered on the Airy disk is dominant in the other three filters. A partial visual detection of the polar dust component can be observed in the F560W filter, however this detection is likely skewed due to the crosshair effect exhibited in this filter. The FWHM increases in synergy with the filter central wavelength and a similar PA as the standard model is seen, but the eccentricity values are lower across all five filters compared to the standard model (Table 3).

$K_{V1}$ shows little evidence of ringing or orange peel effects (except in the F1000W filter), the first Airy ring is well removed, and the ionization bicone is recovered across all filters. A visual detection of the polar dust component can be made in the F560W filter, further evidenced by the higher eccentricity value for this filter. The eccentricity due to the ionization bicone and polar dust is higher than MS$_{V1}$ across all five filters but maintains a similar PA (Table B1). The central

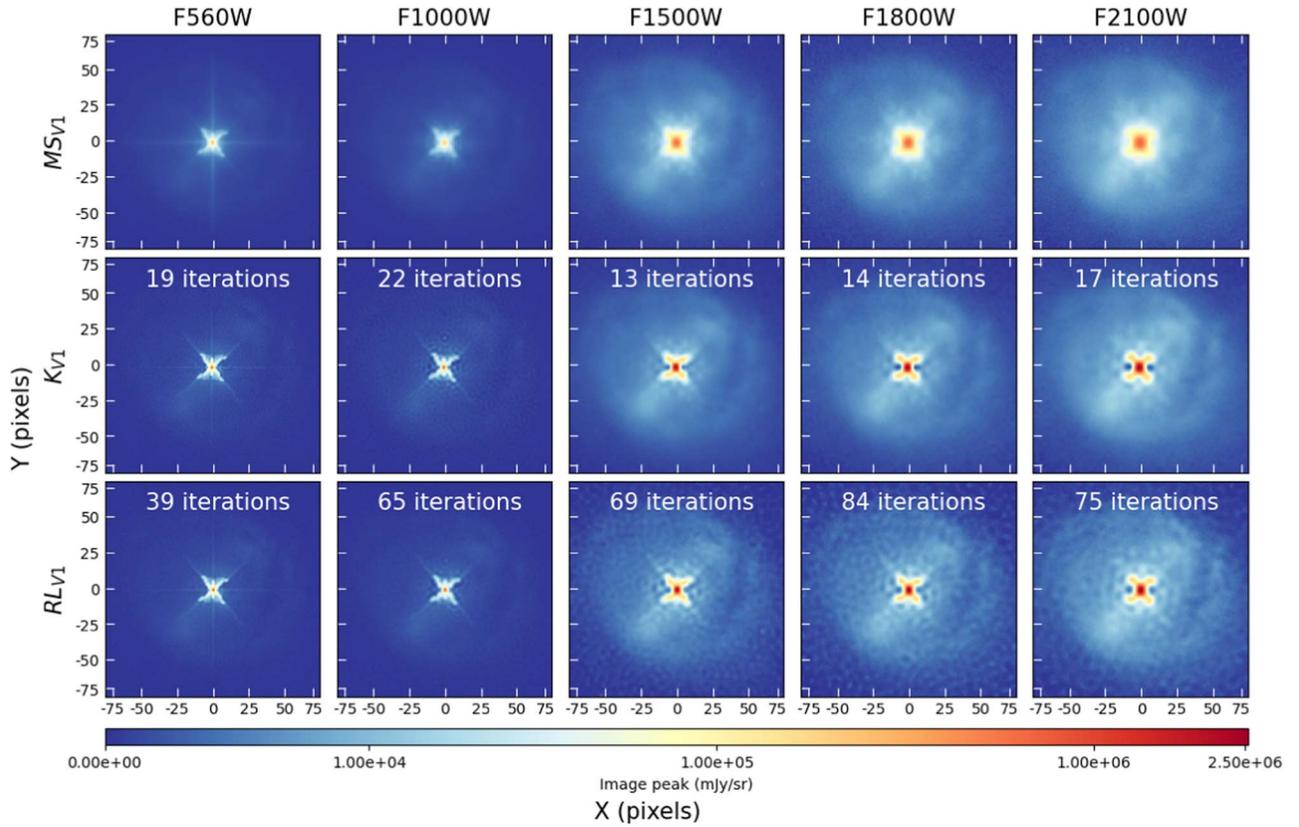

**Figure B1.** Image grid comparison between $MS_{V1}$ (top row), $K_{V1}$ (middle row), and $RL_{V1}$ (bottom row). Each image is displayed log scaled, scaled to the $K_{V1}$ F2100W filter, and shows the central $150 \times 150$ MIRIM pixels.

**Table B1**
Merit Function Results, Position Angle, and Eccentricities for $MS_{V1}$, $K_{V1}$, and $RL_{V1}$

| Filter | F560W | F1000W | F1500W | F1800W | F2100W |
|---|---|---|---|---|---|
| | | | $MS_{V1}$ | | |
| FWHM (pixel) | 4.26 | 5.42 | 7.17 | 8.02 | 9.23 |
| Aperture flux (mJy sr$^{-1}$) | 8.85E+04 | 1.02E+05 | 5.58E+05 | 7.08E+05 | 9.23E+05 |
| PA (°) | 84.36 | 84.0 | 84.81 | 85.30 | 84.89 |
| Eccentricity | 0.36 | 0.37 | 0.35 | 0.34 | 0.32 |
| | | | $K_{V1}$ | | |
| FWHM (pixel) | 1.37 | 2.21 | 3.28 | 3.70 | 4.55 |
| Aperture flux (mJy sr$^{-1}$) | 1.03E+05 | 1.13E+05 | 6.31E+05 | 8.17E+05 | 1.08E+06 |
| PA (°) | 83.79 | 83.99 | 85.43 | 83.40 | 83.97 |
| Eccentricity | 0.47 | 0.47 | 0.49 | 0.48 | 0.47 |
| | | | $RL_{V1}$ | | |
| FWHM (pixel) | 1.84 | 2.45 | 3.49 | 3.80 | 4.58 |
| Aperture flux (mJy sr$^{-1}$) | 1.01E+05 | 1.11E+05 | 6.18E+05 | 7.98E+05 | 1.05E+06 |
| PA (°) | 83.55 | 83.33 | 84.45 | 84.92 | 85.43 |
| Eccentricity | 0.47 | 0.47 | 0.49 | 0.48 | 0.47 |

region's FWHM is reduced by as much as a factor of 3.1, but Δflux shows an increase of up to 17%.

The $R_{V1}$ images show that the first Airy ring is well removed, the ionization bicone is recovered across all filters, and a visual detection of the polar dust component is made in the F560W filter (evidenced again by the higher eccentricity value for this filter). However, like the standard model results (Figure 5), clear evidence of ringing and orange peel effects are observed in the longer-wavelength filter images. The eccentricity due to the ionization bicone and polar dust is again higher than $MS_{V1}$ across all five filters while maintaining a similar PA (Table B1). The central region's FWHM is reduced by as much as a factor of 2.3, and Δflux shows an increase of up to 14%.

Figure B2 shows the merit function (FWHM and aperture flux), eccentricity, and PA values for $MS_{V1}$, $K_{V1}$, and $RL_{V1}$ as a function of simulated central wavelength. As with the standard model, $K_{V1}$ again shows the best improvement in FWHM reduction at all wavelengths, but we note similar improvements in the longer-wavelength filter images as $RL_{V1}$. $K_{V1}$ and $RL_{V1}$ both recovered higher eccentricity values than $MS_{V1}$ and both returned similar PAs as $MS_{V1}$ (with the exception of the $K_{V1}$ F1800W filter image).

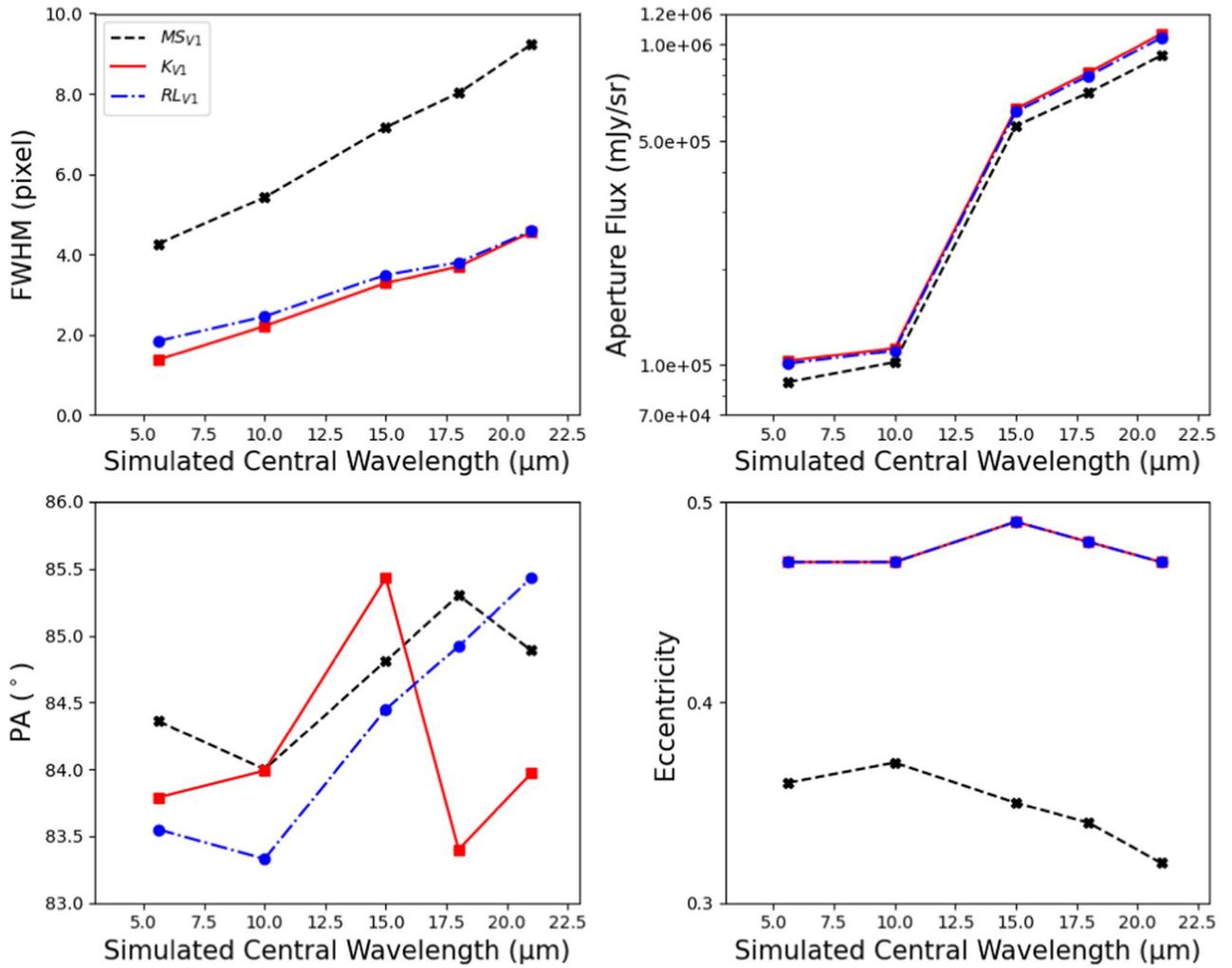

**Figure B2.** FWHM (top left), aperture flux (top right), PA (bottom left), and eccentricity (bottom right) results for $MS_{V1}$, $K_{V1}$, and $RL_{V1}$ as a function of simulated central wavelength.

The $MS_{V2}$ images (Figure B3) show the JWST PSF clearly observed and projected onto the host galaxy in all but the shortest-wavelength filters, with an elongation of the Airy disk noted in the longest-wavelength filter images. A clear visual detection of the polar dust component can be made in the F560W filter, and a partial detection can be made in the F1000W filter (both evidenced by the higher eccentricity values compared to the standard model). The FWHM increases with the filter central wavelength and maintains a similar PA as the standard model (Table 3), but the eccentricity values are higher across all five filters (by as much as 17% in the F560W filter; Table B2).

In the shortest-wavelength filter images of $K_{V2}$ the ionization bicone component is detected, and a clear visual detection of the polar dust component can be made, but evidence of orange peel effects persist in the F1000W filter. The longer-wavelength filters again exhibit little evidence of ringing or orange peel effects, however the ionization bicone component is not well recovered. Instead, a remnant of the first Airy ring is present, and a feature similar in morphology to the polar dust component visually dominates the central region at these wavelengths (evidenced by the similar eccentricity and PA values across all five filters for $K_{V2}$; Table B2). The central region's FWHM is reduced by as much as a factor of 2.5, but $\Delta$flux shows an increase of up to 16%.

Similar to $K_{V2}$, the $RL_{V2}$ observations show a clear detection of the polar dust component in the shortest-wavelength filter images, but the ionization bicone component is again not well recovered in the longest-wavelength filter images. Instead, a residual of the first Airy ring is present, and a feature similar in morphology to the polar dust component again visually dominants the central region at these wavelengths. The central region's FWHM is reduced by as much as a factor of 2.2, and $\Delta$flux shows an increase of up to 14% (Table B2).

Figure B4 shows the merit function (FWHM and aperture flux), eccentricity, and PA values between $MS_{V2}$, $K_{V2}$, and $RL_{V2}$ as a function of simulated central wavelength. Following the results of the standard model and $K_{V1}$, $K_{V2}$ again shows the best improvement in FWHM reduction at all wavelengths with similar improvements in the longer-wavelength filter images as $RL_{V2}$. $K_{V2}$ and $RL_{V2}$ again both recovered higher eccentricity values than $MS_{V2}$ (with $RL_{V2}$ performing marginally better in the longer-wavelength filters than $K_{V2}$) and both returned similar PAs as $MS_{V2}$.

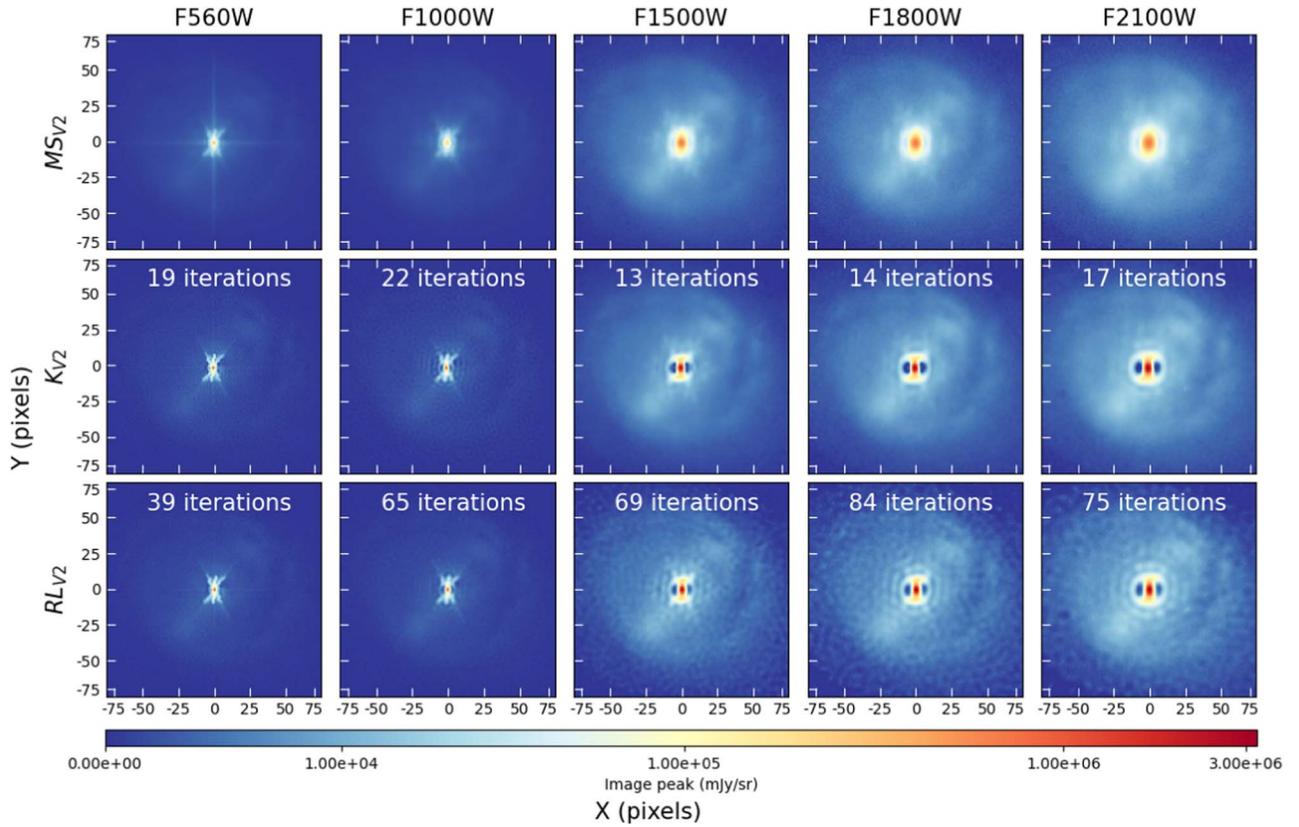

**Figure B3.** Same as Figure B1 but showing $MS_{V2}$ (top row), $K_{V2}$ (middle row), and $RL_{V2}$ (bottom row).

Table B2
Merit Function Results, Position Angle, and Eccentricities for $MS_{V2}$, $K_{V2}$, and $RL_{V2}$

| Filter | F560W | F1000W | F1500W | F1800W | F2100W |
|---|---|---|---|---|---|
| | | | $MS_{V2}$ | | |
| FWHM (pixel) | 3.14 | 4.56 | 6.47 | 7.41 | 8.64 |
| Aperture flux (mJy sr$^{-1}$) | 8.85E+04 | 1.02E+05 | 5.58E+05 | 7.08E+05 | 9.23E+05 |
| PA (°) | 84.25 | 84.07 | 84.44 | 84.03 | 83.84 |
| Eccentricity | 0.55 | 0.56 | 0.54 | 0.51 | 0.47 |
| | | | $K_{V2}$ | | |
| FWHM (pixel) | 1.26 | 1.93 | 2.86 | 3.22 | 3.99 |
| Aperture flux (mJy sr$^{-1}$) | 1.02E+05 | 1.12E+05 | 6.27E+05 | 8.18E+05 | 1.07E+06 |
| PA (°) | 84.33 | 84.49 | 84.65 | 83.83 | 83.64 |
| Eccentricity | 0.67 | 0.67 | 0.67 | 0.66 | 0.64 |
| | | | $RL_{V2}$ | | |
| FWHM (pixel) | 1.86 | 2.33 | 3.10 | 3.31 | 4.06 |
| Aperture flux (mJy sr$^{-1}$) | 1.01E+05 | 1.11E+05 | 6.16E+05 | 7.96E+05 | 1.05E+06 |
| PA (°) | 84.21 | 84.34 | 84.37 | 84.38 | 83.91 |
| Eccentricity | 0.67 | 0.67 | 0.68 | 0.67 | 0.66 |

For both model variations, Kraken and Richardson–Lucy performed very similarly to the standard model, following similar trends in terms of merit function results and image quality improvement. In terms of FWHM reduction for both model variations Kraken performed the best in the shortest-wavelength filters, while Kraken and Richardson–Lucy performed similarly in the longest-wavelength filters (with Kraken performing the best). In terms of flux conservation, Richardson–Lucy was the most flux conservative for both model variations but both algorithms performed very similarly across all five filters. In terms of image quality, both algorithms significantly improved the image quality across all five filters, but Kraken introduced the fewest ringing and orange peel effects (especially in the longest-wavelength filters). That these trends observed were similar across all five filters between the standard model and the two model variations gives confidence to the validity of our deconvolution methodology and its application to different configurations of the standard model.

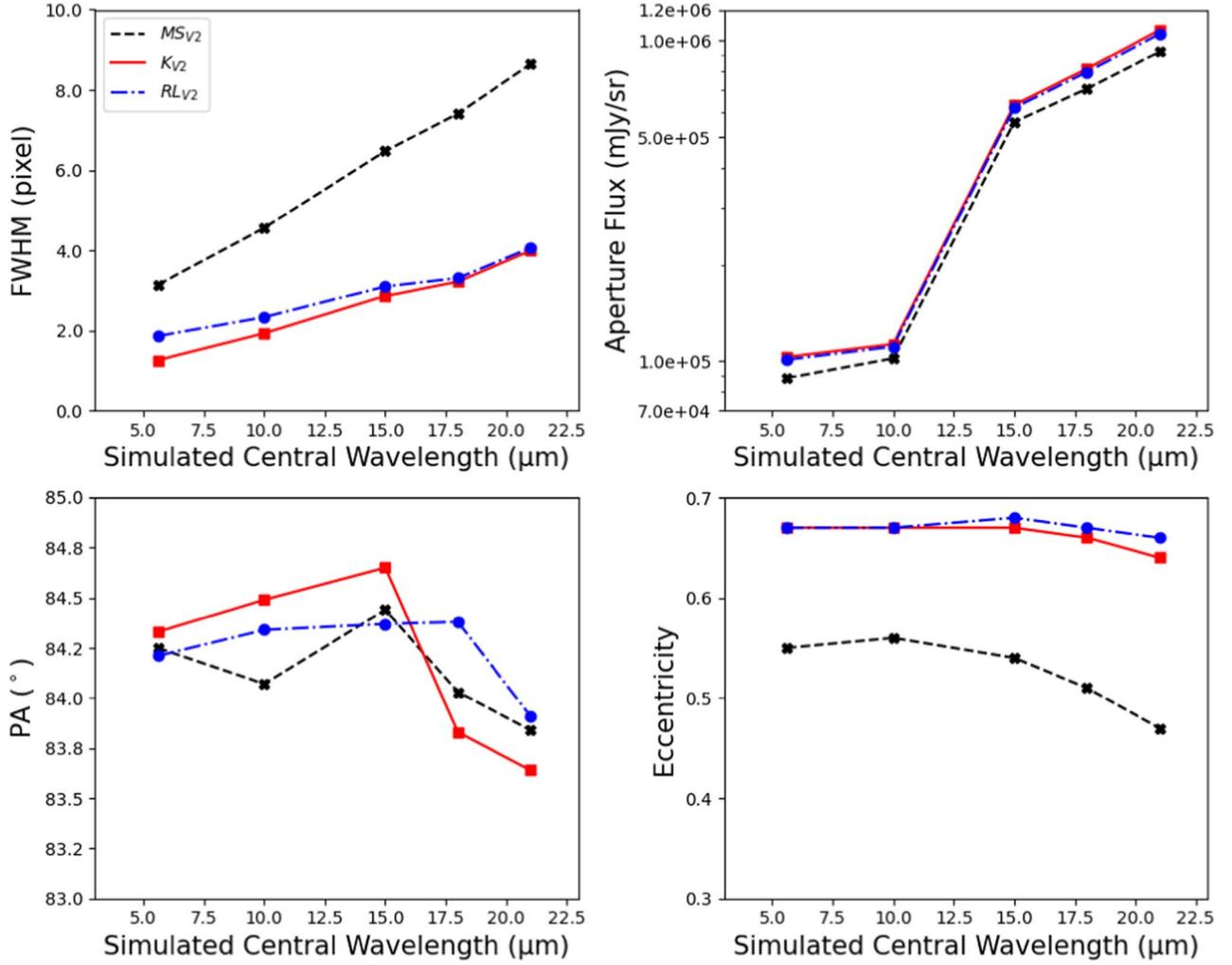

**Figure B4.** Same as Figure B2, but for $MS_{V2}$, $K_{V2}$, and $RL_{V2}$.

### B.2. Excluding Model Components

To explore Kraken's performance with different models, we generated two variations of the standard model where a single component was sequentially excluded from the MIRISim input (Section 2.2). Each model maintained the same physical extent and intensity of the model components used in the standard model (Table 1) but either the polar dust component ($V_3$) or the ionization bicone component ($V_4$) were excluded. MS images for $V_{3-4}$ were generated ($MS_{V3-4}$) in the mid to lowest spatial resolution filters (F1500W, F1800W, and F2100W) then Kraken deconvolved ($D_{V3-4}$) to the same number of iterations as the Kraken deconvolved standard model ($D_{AGN}$; Table 3) for each filter. The filter choice was motivated by our desire to recover the polar dust and bicone component at wavelengths where these components were not readily detectable in the MS images (see Figure 5).

Figure B5 shows $D_{AGN}$ and $D_{V3-4}$ for each filter, where the first Airy ring and the ionization bicone component are clearly recovered in $D_{AGN}$ and $D_{V3}$. $D_{V4}$ clearly recovers an elongation similar in morphology to the polar dust component and a remnant of the first Airy ring is present in all filters (similar to the results of $K_{V2}$; Figure B3). Figure B6 shows the merit function (FWHM and aperture flux), eccentricity, and PA values between $D_{AGN}$ and $D_{V3-4}$ as a function of simulated central wavelength. Each model performed similarly in terms of merit functions, eccentricity, and PAs ($\Delta < 1\%$) but $D_{V4}$ exhibited higher eccentricity values across all filters (Table B3).

### B.3. Faint Source Detection

An important benefit of any deconvolution algorithm is the ability to recover a faint source (FS) near the PSF. To explore Kraken's ability to recover FSs, we generated variations of the standard model (maintaining the same physical extent and intensity of each model component) but adding an FS to the model ($V_5$). The FS was modeled as a 2D Gaussian (FWHM = 3.0 pixels) with the total integrated counts set to 75:1 compared to the central region PSF (i.e., fainter than the bicone component but brighter than the host galaxy). The FS was initially placed such that the FS x-centroid position was offset by 1 pixel from the apex of the ionization bicone component of the standard model. MS images of V5 were generated ($MS_{V5}$; Section 2.2) in the lowest spatial resolution filter (F2100W, i.e., the PSF with the largest Airy disk/ring; Figure 1) then Kraken deconvolved ($D_{V5}$) to the same number of iterations as the Kraken deconvolved standard model (Table 3). The x-offset of the FS was then increased by a single pixel and the process repeated.

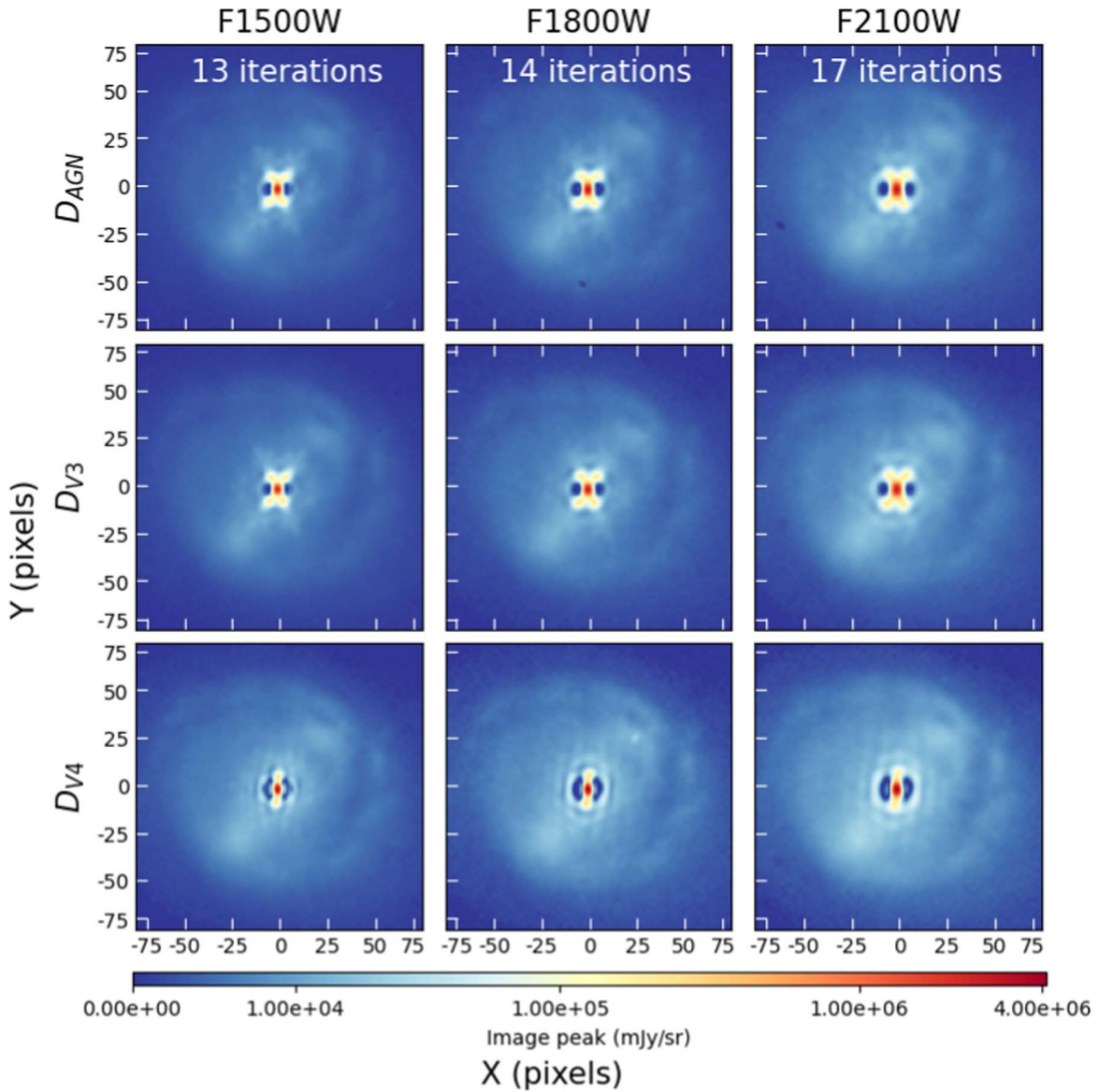

**Figure B5.** Kraken deconvolution imaging results for $D_{AGN}$ (top row), $D_{V3}$ (middle row), and $D_{V4}$ (bottom row). Each image is displayed log scaled, scaled to the F2100W for each row, and shows the central 150 × 150 MIRIM pixels. The number of iterations for each column of images is displayed as an inset in the $D_{AGN}$ images.

Figure B7 shows the $MS_{V5}$ and $D_{V5}$ images where the FS was first detected. DAOStarFinder[48] (Stetson 1987) was used to detect where the FS was first detected to a 3σ certainty for $MS_{V5}$ (9 pixels, 0″.99, 187.2 pc) and $D_{V5}$ (6 pixels, 0″.66, 124.8 pc). $D_{V5}$ was able to recover a faint object with a Δ of 3 pixels (0″.33, 62.4 pc) closer to the central region compared to $MS_{V5}$.

*B.4. Deconvolution Performance versus Signal-to-noise Ratio*

The performance and resulting fidelity of any deconvolution algorithm is critically dependent on the S/N of the image being deconvolved (Davies & Kasper 2012). When the S/N of the image to be deconvolved is high, the deconvolution algorithm can easily converge, but for lower-S/N data the convergence can be difficult to reach or improperly recovers the image.

---

[48] https://photutils.readthedocs.io/en/stable/api/photutils.detection.DAOStarFinder.html

Further, if the deconvolution algorithm improperly handles noise, such noise may be taken as a true signal and amplified (McNeil & Moody 2005). Magain et al. (1998) describe these effects for Richardson–Lucy deconvolution.

As flux is redistributed during deconvolution iterations, an increased flux measurement in small-sized apertures near to a point source is found, demonstrating convergence of the algorithm (Section 3.2). This effect can be used to confirm the effectiveness of the deconvolution in images with low S/Ns. To explore Kraken's performance at lower S/Ns we generated variations of the standard model, sequentially reducing the total integrated model counts in 10 steps by 1000 counts each step, reducing the input signal of the model ($V_6$) being propagated through MIRISim but at a constant noise level. MS images were produced for each variation of $V_6$ ($MS_{V6}$, Section 2.2) in the lowest spatial resolution filter (F2100W) then Kraken deconvolved ($D_{V6}$) to the same number of iterations as the Kraken deconvolved standard model (Table 2). The F2100W filter was selected to test Kraken's performance at recovering model components in the lowest spatial resolution filter image. For the noise per pixel we used an average value of the rms of four square 10 × 10 pixel apertures (one in each corner of the array). The signal was first measured within a circular aperture at the centroid that encircled 95% of the flux (AS$_{\mathrm{total}}$, 174.8 pixel diameter). Next, the signal dominated by the central region, polar dust, and bicone component was determined within a circular aperture (AS$_{\mathrm{small}}$, 36 pixel diameter, Section 3.2), placed at the centroid position.

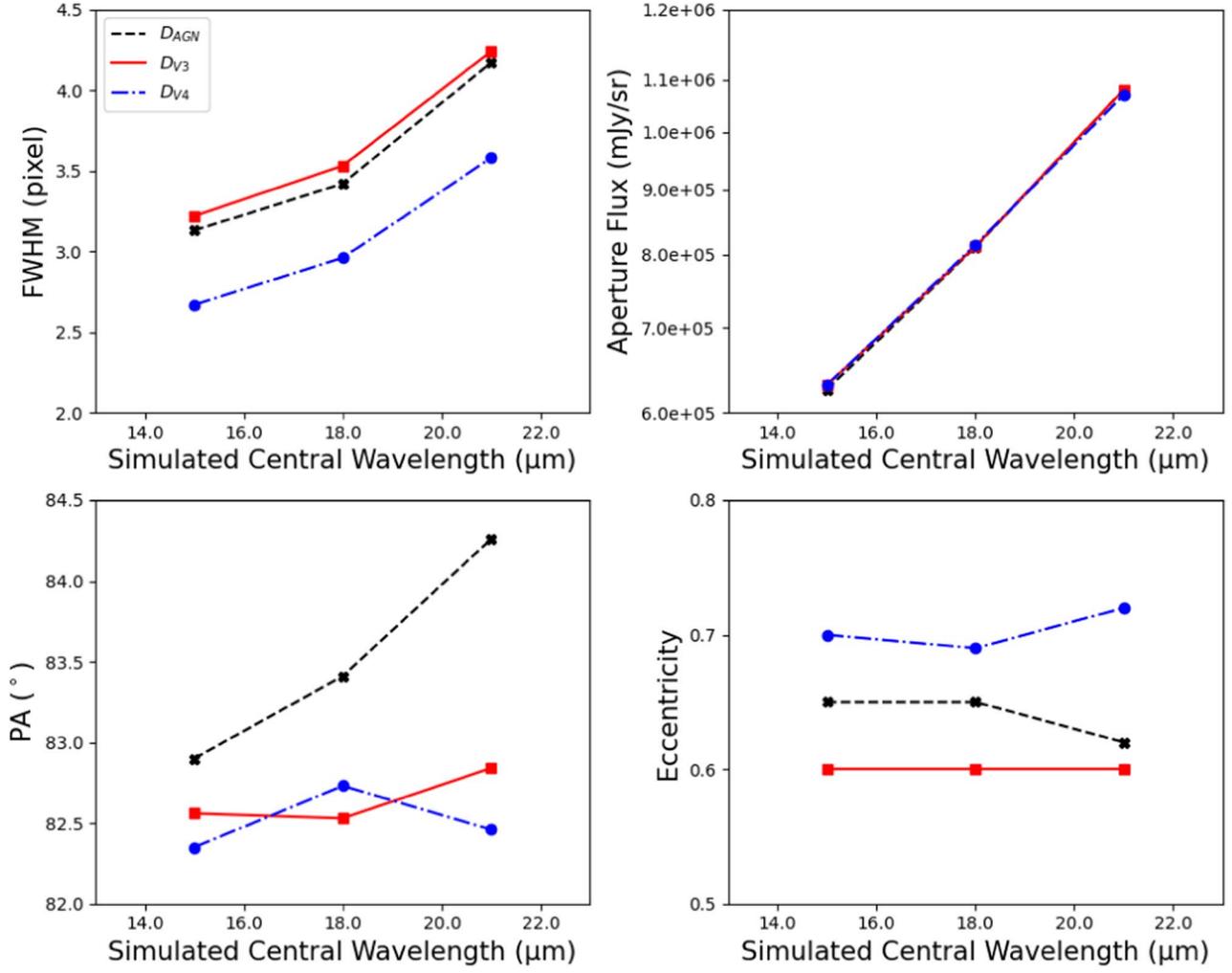

**Figure B6.** Same as Figure B2, but for $D_{\mathrm{AGN}}$, $D_{V3}$, and $D_{V4}$.

**Table B3**
Merit Function Results, Position Angle, and Eccentricities for $D_{V3}$ and $D_{V4}$

| Filter | F1500W | F1800W | F2100W |
|---|---|---|---|
| | $D_{V3}$ | | |
| FWHM (pixel) | 3.22 | 3.53 | 4.24 |
| Aperture flux (mJy sr$^{-1}$) | 6.31E+05 | 8.12E+05 | 1.08E+06 |
| PA (°) | 82.56 | 82.53 | 82.84 |
| Eccentricity | 0.60 | 0.60 | 0.60 |
| | $D_{V4}$ | | |
| FWHM (pixel) | 2.67 | 2.96 | 3.58 |
| Aperture flux (mJy sr$^{-1}$) | 6.31E+05 | 8.14E+05 | 1.07E+06 |
| PA (°) | 82.35 | 82.73 | 82.46 |
| Eccentricity | 0.70 | 0.69 | 0.72 |

Table B4 shows the S/N, noise per pixel, and aperture signal measurements for each variation of MS$_{V6}$ and $D_{V6}$. Figure B8 shows the MS$_{V6}$ and $D_{V6}$ aperture signal measurements as a function of S/N. AS$_{\mathrm{total}}$ again demonstrates that Kraken deconvolution is flux consistent to $\lesssim$5% at all S/Ns for a large aperture. However, AS$_{\mathrm{small}}$ shows a significant flux difference (up to 16.1%) between MS$_{V6}$ and $D_{V6}$, consistent with flux redistribution increasing the flux in a small aperture and with the standard model results (Table 2). At S/N = 370 the difference in AS$_{\mathrm{small}}$ between MS$_{V6}$ and $D_{V6}$ decreased by only ∼6.5%, demonstrating deconvolution is no longer effective at S/Ns of around this value or below.

Figures B9 and B10 show the MS$_{V6}$ and $D_{V6}$ results for the high to low S/N images, with the MS$_{V6}$ S/N and $D_{V6}$ recovered S/N values given. For the higher S/N MS$_{V6}$ images, the galaxy component is clearly recovered and the complex JWST PSF is clearly observed. The corresponding $D_{V6}$ image shows a substantial improvement in the S/N (Table B4) compared to MS$_{V6}$, with the Airy ring well removed, and a clear recovery of the ionization bicone and galaxy component. However, at S/N $\lesssim$ 370 the galaxy component is no longer visually detected in the MS$_{V6}$ image and $D_{V6}$ images, and the ionization bicone cannot be well distinguished in the $D_{V6}$ image. Where the S/N $\lesssim$ 370 visually showing the lack of devolution effectiveness is consistent with the results shown in Figure B8.

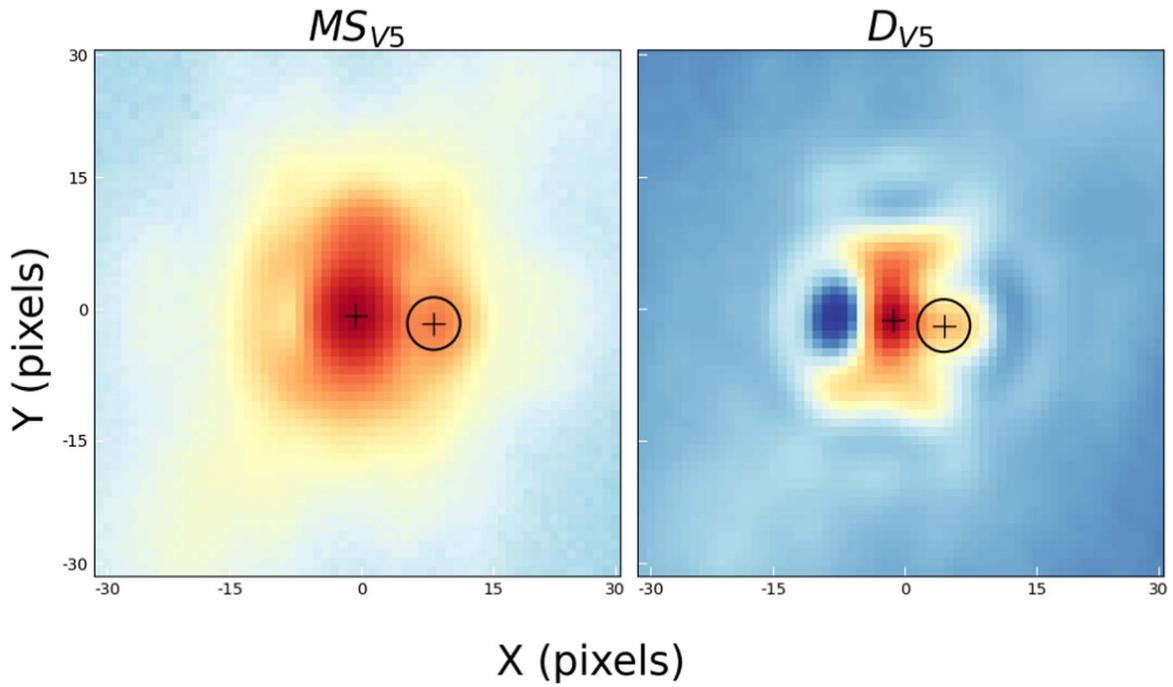

**Figure B7.** FS detection comparisons between $MS_{V5}$ (right) and $D_{V5}$ (left). Each image is log scaled and shows the central 60 × 60 MIRIM pixels. For both images the centroid locations of the nucleus and FS are marked with a "+" and the FS detected with a 3σ certainty is highlighted with a black aperture.

**Table B4**
Signal-to-noise Ratio, Noise per Pixel, and Aperture Signal Results for $MS_{V6}$ and $D_{V6}$ for Each Signal-to-noise Ratio Level

| $MS_{V6}$ | | | | $D_{V6}$ | | | | | |
|---|---|---|---|---|---|---|---|---|---|
| $AS_{small}$ (mJy sr$^{-1}$) | $AS_{total}$ (mJy sr$^{-1}$) | Noise per pixel (mJy sr$^{-1}$) | S/N$_{total}$ | $AS_{small}$ (mJy sr$^{-1}$) | $\Delta_{small}$ (%) | $AS_{total}$ (mJy sr$^{-1}$) | $\Delta_{total}$ (%) | Noise per pixel (mJy sr$^{-1}$) | S/N$_{total}$ |
| 9.78E+05 | 1.97E+06 | 2.52E+00 | 5.05E+03 | 1.14E+06 | 16.12 | 2.03E+06 | 3.06 | 9.04E-01 | 1.45E+04 |
| 8.30E+05 | 1.68E+06 | 2.56E+00 | 4.23E+03 | 9.63E+05 | 16.09 | 1.72E+06 | 2.63 | 8.73E-01 | 1.28E+04 |
| 6.96E+05 | 1.40E+06 | 2.49E+00 | 3.64E+03 | 8.08E+05 | 16.04 | 1.43E+06 | 2.03 | 9.03E-01 | 1.02E+04 |
| 5.61E+05 | 1.13E+06 | 2.59E+00 | 2.82E+03 | 6.50E+05 | 15.98 | 1.14E+06 | 1.17 | 9.68E-01 | 7.60E+03 |
| 4.23E+05 | 8.48E+05 | 2.56E+00 | 2.15E+03 | 4.90E+05 | 15.83 | 8.47E+05 | 0.22 | 7.88E-01 | 6.93E+03 |
| 2.85E+05 | 5.69E+05 | 2.55E+00 | 1.45E+03 | 3.29E+05 | 15.62 | 5.53E+05 | 2.77 | 8.88E-01 | 4.02E+03 |
| 1.44E+05 | 2.85E+05 | 2.67E+00 | 7.01E+02 | 1.64E+05 | 14.13 | 2.70E+05 | 5.38 | 9.49E-01 | 1.75E+03 |
| 7.54E+04 | 1.69E+05 | 2.48E+00 | 5.24E+02 | 8.40E+04 | 11.43 | 1.61E+05 | 4.61 | 8.57E-01 | 1.21E+03 |
| 7.20E+04 | 1.43E+05 | 2.56E+00 | 3.70E+02 | 7.67E+04 | 6.48 | 1.40E+05 | 1.99 | 8.57E-01 | 1.06E+03 |
| 4.82E+04 | 9.07E+04 | 2.48E+00 | 2.43E+02 | 5.04E+04 | 4.50 | 9.33E+04 | 2.87 | 8.57E-01 | 7.03E+02 |

**Note.** The S/N reported are for $AS_{total}$. The $\Delta_{total}$ and $\Delta_{small}$ are the total and small aperture differences between $MS_{V6}$ and $D_{V6}$ for each S/N level.

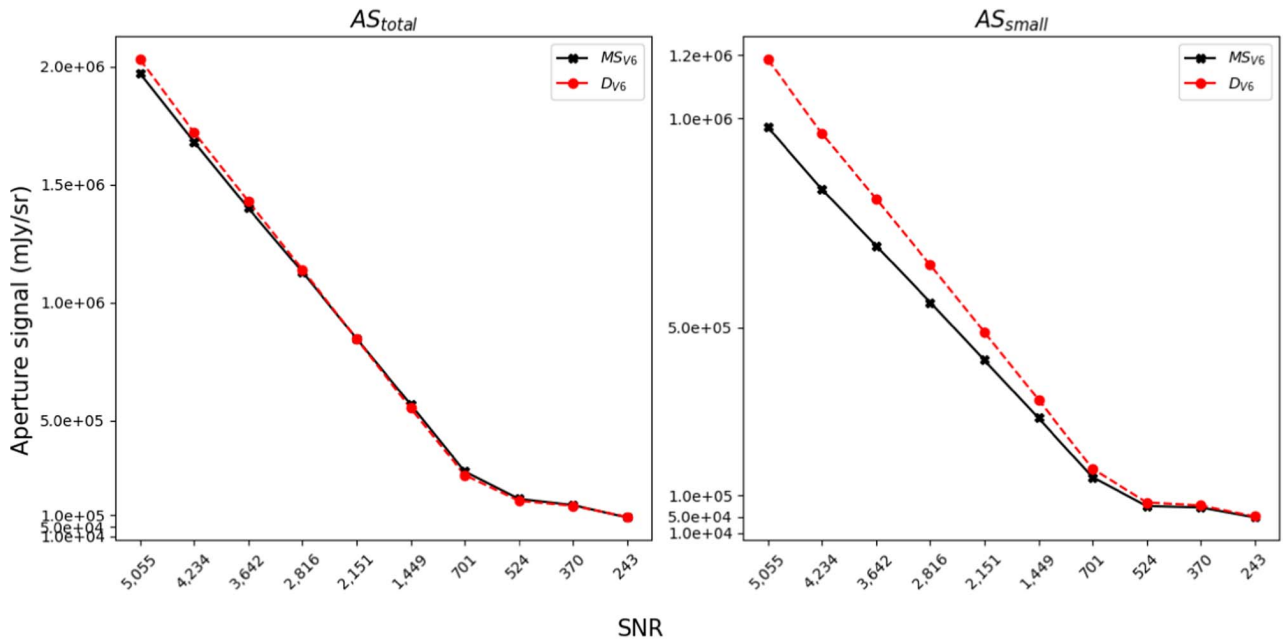

**Figure B8.** $MS_{V6}$ and $D_{V6}$ aperture signal measurements for $AS_{total}$ (left) and $AS_{small}$ (right) as a function of S/N.

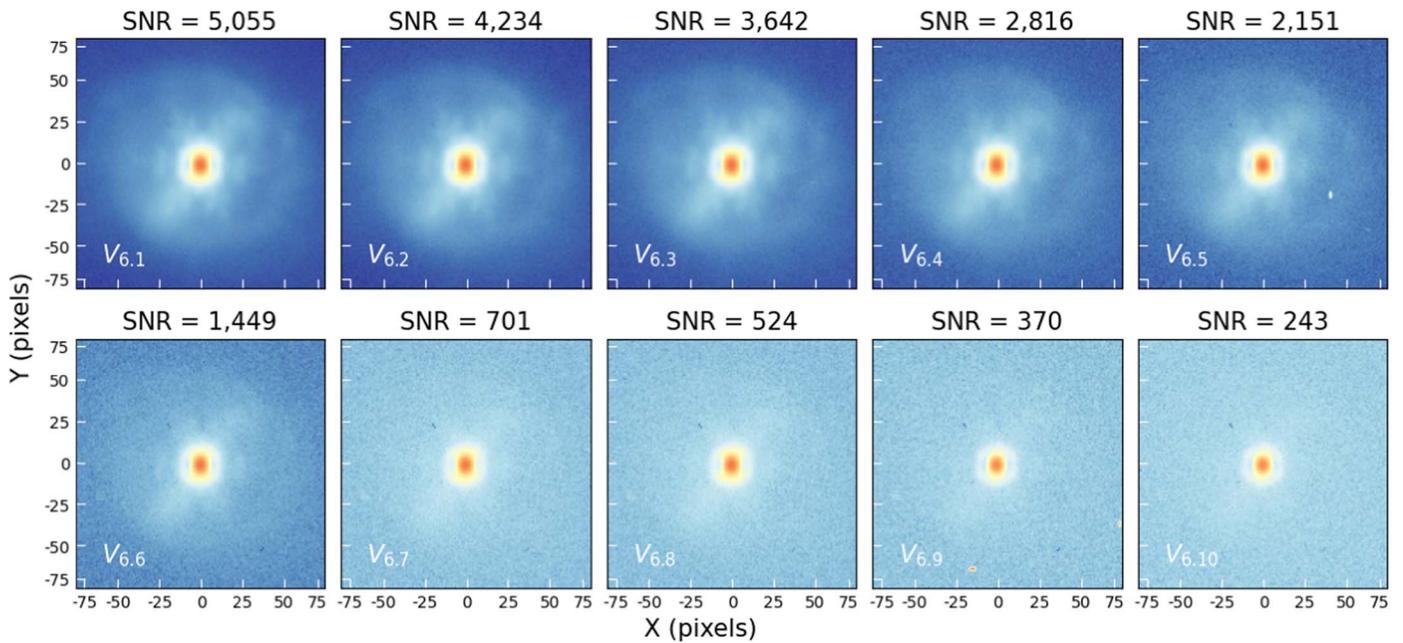

**Figure B9.** High to low S/N $MS_{V6}$ images, where the S/N for each image is given in the title. Each image is displayed log scaled, scaled to the corresponding $D_{V6}$ image peak (Figure B10), and shows the central 150 × 150 MIRIM pixels. The input model variation ($V_6$) number is given as an inset in each image.

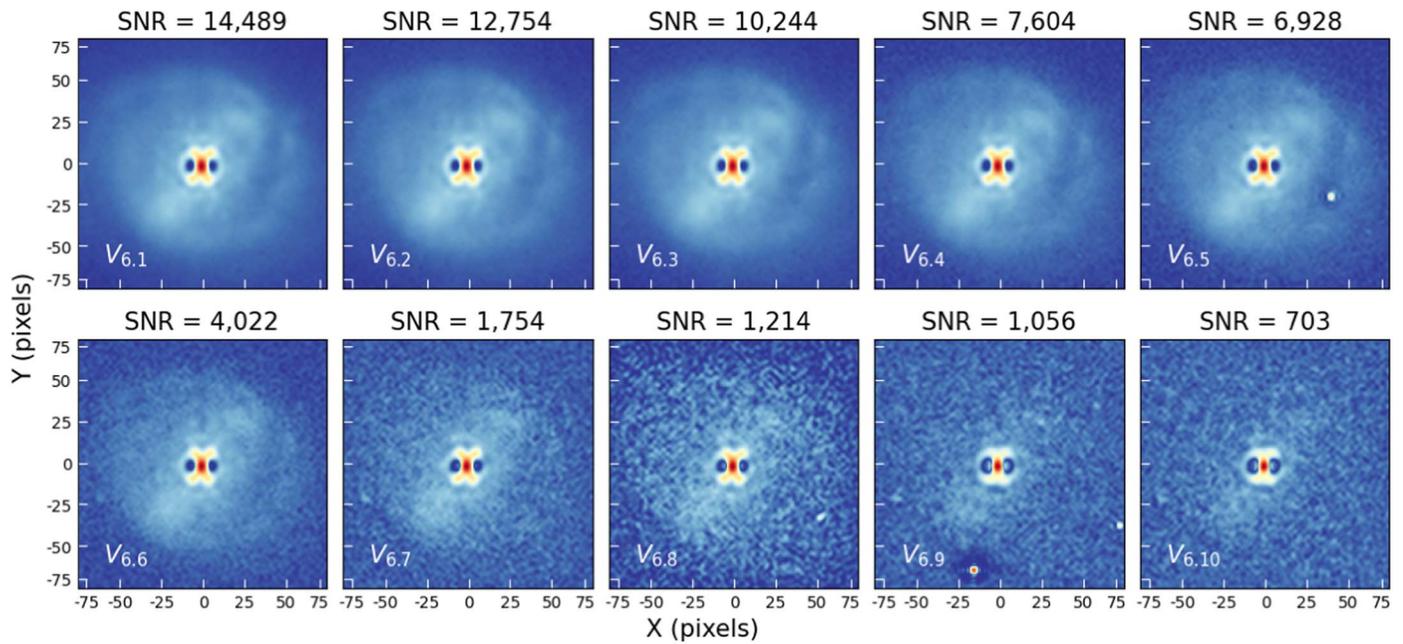

**Figure B10.** Same as Figure B9, but for the $D_{V6}$ images.

## ORCID iDs

M. T. Leist ⓘ https://orcid.org/0000-0003-4975-2046
C. Packham ⓘ https://orcid.org/0000-0001-7827-5758
D. J. V. Rosario ⓘ https://orcid.org/0000-0002-0001-3587
D. A. Hope ⓘ https://orcid.org/0000-0002-7957-7018
A. Alonso-Herrero ⓘ https://orcid.org/0000-0001-6794-2519
E. K. S. Hicks ⓘ https://orcid.org/0000-0002-4457-5733
L. Zhang ⓘ https://orcid.org/0000-0003-4937-9077
R. Davies ⓘ https://orcid.org/0000-0003-4949-7217
T. Díaz-Santos ⓘ https://orcid.org/0000-0003-0699-6083
O. González-Martín ⓘ https://orcid.org/0000-0002-2356-8358
E. Bellocchi ⓘ https://orcid.org/0000-0001-9791-4228
P. G. Boorman ⓘ https://orcid.org/0000-0001-9379-4716
F. Combes ⓘ https://orcid.org/0000-0003-2658-7893
I. García-Bernete ⓘ https://orcid.org/0000-0002-9627-5281
S. García-Burillo ⓘ https://orcid.org/0000-0003-0444-6897
B. García-Lorenzo ⓘ https://orcid.org/0000-0002-7228-7173
K. Ichikawa ⓘ https://orcid.org/0000-0002-4377-903X
M. Imanishi ⓘ https://orcid.org/0000-0001-6186-8792
S. M. Jefferies ⓘ https://orcid.org/0000-0002-9580-5615
Á. Labiano ⓘ https://orcid.org/0000-0002-0690-8824
N. A. Levenson ⓘ https://orcid.org/0000-0003-4209-639X
R. Nikutta ⓘ https://orcid.org/0000-0002-7052-6900
M. Pereira-Santaella ⓘ https://orcid.org/0000-0002-4005-9619
C. Ramos Almeida ⓘ https://orcid.org/0000-0001-8353-649X
C. Ricci ⓘ https://orcid.org/0000-0001-5231-2645
D. Rigopoulou ⓘ https://orcid.org/0000-0001-6854-7545
W. Schaefer ⓘ https://orcid.org/0009-0008-9534-4661
M. Stalevski ⓘ https://orcid.org/0000-0001-5146-8330
L. Fuller ⓘ https://orcid.org/0000-0003-4809-6147
T. Izumi ⓘ https://orcid.org/0000-0001-9452-0813
T. Shimizu ⓘ https://orcid.org/0000-0002-2125-4670